\documentclass[
reprint,
superscriptaddress,
amsmath,amssymb,
aps,
prb,
floatfix
]{revtex4-1}

\usepackage{color}
\usepackage{graphicx}
\usepackage{dcolumn}
\usepackage[normalem]{ulem}
\usepackage{bm}
\usepackage{mathtools}
\usepackage{bbold}

\def\k{{\bf k}}
\def\r{{\bf r}}

\begin{document}

\preprint{}

\title{Correlations Among STM Observables in Disordered Unconventional Superconductors}

\author{Miguel Antonio Sulangi}
\affiliation{
	Department of Physics, University of Florida, Gainesville, FL 32611, USA
}

\author{W. A. Atkinson}
\affiliation{
	Department of Physics and Astronomy, Trent University, Peterborough, Ontario K9L 0G2, Canada
}

\author{P. J. Hirschfeld}
\affiliation{
	Department of Physics, University of Florida, Gainesville, FL 32611, USA
}

\date{\today}

\begin{abstract}
New developments in scanning tunneling spectroscopy now allow for the spatially resolved measurement of the Josephson critical current $I_c$ between a tip and a superconducting sample, a nearly direct measurement of the true superconducting order parameter. However, it is unclear how these $I_c$ measurements are correlated with previous estimates of the spectral gap taken from differential conductance measurements. In particular, recent such experiments on an iron-based superconductor found almost no correlation between $I_c$ and the spectral gap obtained from differential conductance $g=dI/dV$ spectra, reporting instead a more significant correlation between $I_c$ and the the coherence-peak height. Here we point out that the correlation---or the lack thereof---between these various quantities can be naturally explained by the effect of disorder on unconventional superconductivity. Using large scale numerical simulations of a BCS $d$-wave pair Hamiltonian with many-impurity potentials, we observe that ``substitutional'' disorder models with weak pointlike impurities lead to a situation in which the true superconducting order parameter and $I_c$ are both uncorrelated with the spectral gap from $dI/dV$ measurements and highly correlated with the coherence-peak heights.  The  underlying mechanism appears to be the disorder-induced transfer of spectral weight away from the coherence peaks. On the other hand, smooth impurity potentials with a length scale larger than the lattice constant lead to a large positive correlation between the true superconducting order parameter and the spectral gap, in addition to a large correlation between the order parameter and the coherence-peak height. We discuss the applicability of our results to recent Josephson scanning tunneling spectroscopy experiments on iron-based and cuprate high-temperature superconductors.
\end{abstract}

\maketitle

\section{\label{sec:level1}Introduction}

A good deal of what we presently know about cuprate and other unconventional superconductors is due to scanning tunneling spectroscopy (STS), which over the past several decades has uncovered a panoply of exotic phenomena in these materials.\cite{schmidt2011electronic} Thanks  to advances in experimental techniques, the ability to resolve in great detail the spatial features of these materials has shown that at least some of the cuprates are strongly inhomogeneous. Manifestations of the inhomogeneous character of the cuprates as seen by STS include quasiparticle scattering interference\cite{hoffman2002imaging,mcelroy2003relating,hanaguri2007quasiparticle,kohsaka2008cooper,fujita2014simultaneous,he2014fermi}, charge order\cite{howald2003coexistence, howald2003periodic,vershinin2004local,hanaguri2004checkerboard,kohsaka2007intrinsic,lawler2010intra}, and a strongly inhomogeneous superconducting gap\cite{howald2001inherent,lang2002imaging,fang2004periodic,mcelroy2005atomic,fang2006gap,gomes2007visualizing,pasupathy2008electronic,alldredge2008evolution}. The main tool of  STS is the measurement of the differential conductance, which is proportional to the local density of states (LDOS) and thus reveals much about the electronic spectral properties of these materials.

Recent technical advances have resulted in a variant of the experimental technique called Josephson scanning tunneling spectroscopy (JSTS), which makes use of quantum-mechanical tunneling of Cooper pairs between a superconducting tip and the sample to map the spatial variations of the critical current.\cite{smakov2001josephson,naaman2001fluctuation,rodrigo2004use,proslier2006probing} The technique has recently been applied to the underdoped Bi$_2$Sr$_2$CaCu$_2$O$_{8+\delta}$ (BSCCO) and to the iron-based superconductor FeTe$_{0.55}$Se$_{0.45}$ (FeTeSe), both of which were  shown to exhibit strongly inhomogeneous superconducting order\cite{cho2019strongly}. In addition,  BSCCO exhibits an eight-unit cell modulation of the superconducting wave function\cite{hamidian2016detection} (pair density wave, or PDW).

JSTS is similar to STS, but with superconducting tips instead of metallic ones so that the tunneling process can be understood as that of a very small superconductor-insulator-superconductor junction. The critical current is observable at very small bias voltages, reflecting Cooper-pair tunneling between the tip and the sample.\cite{ivanchenko1969josephson,ingold1994cooper} This is a key probe of the ground-state properties of the superconducting condensate---in particular, of the superconducting order parameter. The measurement of the local critical current $I_c$ from JSTS has been shown in theoretical work, and confirmed here, to be an excellent proxy for the superconducting order parameter, so one may take the recent results from JSTS experiments to be an accurate picture of the strongly inhomogeneous nature of the superconductivity in these materials.\cite{graham2017imaging,graham2019josephson}

It is natural to compare results from differential conductance and JSTS, because prior to the advent of JSTS, the spectral gap maps $\Omega(\r)$ obtained from $dI/dV$ measurements were often assumed to represent the spatially resolved superconducting order parameter $\Delta_{\k}(\r)$, which itself is not directly observable. Intuitively, there is no reason to suspect that there should be a discrepancy between the spectral gap and the $I_c$ maps, since they should both reflect the underlying superconducting order parameter. For example, while the 8$a_0$ critical current oscillations reported in BSCCO JSTS were not initially observed in $g(\r)$ maps \cite{hamidian2016detection}, their existence was established recently\cite{Du2020}.  However, the {\it local} correlations between the two are not known. Understanding the correlations among these observables may be the clue to identifying the type of disorder present in the BSCCO system, and thereby help to isolate the intrinsic physics of the underdoped cuprates.

A second indication that the correlation between the local critical current and the spectral gap was weak was discovered in the aforementioned JSTS measurement on FeTeSe, which showed that the spectral gap was almost completely uncorrelated with the quantity $I^2_c R^2_N$, which should be proportional to the square of the superconducting order parameter ($R_N$ is the normal-state junction resistance).\cite{cho2019strongly} The discrepancy between $I_c$ and the spectral gap  in FeTeSe was interpreted as equivalent to that between the superfluid density $\rho_s$ and the superconducting order parameter---that is, $I^2_c R^2_N$ was interpreted  as a proxy for $\rho_s$. However, there is no reason for the two quantities to be proportional to one another, since the latter is determined strictly by normal-state quantities in BCS theory, while the former is related to the order parameter via the Ambegaokar-Baratoff relation.\cite{ambegaokar1963tunneling} Hence the observed discrepancy between the spectral gap and $I^2_c R^2_N$ cannot be explained by interpreting the latter quantity as the superfluid density. It was also found in the FeTeSe experiment that the $I^2_c R^2_N$ maps were instead much more correlated with the coherence-peak height, but it was unclear as to why this was the case. In any case, the evidence from the first few sets of JSTS experiments is clear: the spectral gap from $dI/dV$ measurements and $I_c$ are not necessarily correlated with each other. Why this is the case is not presently understood.

In this paper we set out to explain this conundrum by revisiting a very well-trodden path: disorder in $d$-wave superconductors.\cite{gorkov1985defects,hirschfeld1988consequences,lee1993localized,franz1996impurity,franz1997critical,atkinson2000details,atkinson2000gap,durst2000impurity,ghosal2000spatial,durst2002microwave,nunner2005microwave,nunner2005dopant,chakraborty2014fate,sulangi2017revisiting,sulangi2018quasiparticle,Roemer2018,lee2018optical,lee2020low,li2021superconductor} We demonstrate that the lack of correlation between the spectral gap and the true order parameter, as probed by $I_c$, is a  natural consequence of disorder. We illustrate various scenarios in which this absence of correlation between ostensibly similar quantities arises for some models of disorder, but not others. The models of disorder we study in detail are weak pointlike scatterers, binary-alloy disorder, and smooth screened Coulomb-potential disorder. We find that when disorder is pointlike in nature, the correlation between the spectral gap and the true order parameter (which correlates very strongly with the critical current) is typically quite weak. On the other hand, when the disorder potential is extended, these two quantities become much more strongly correlated with each other. By obtaining these correlation coefficients for different disorder strengths, we identify disorder regimes that appear to describe BSCCO and FeTeSe well. 

We also find a strong correlation between the order parameter and the coherence-peak height. We illustrate this mechanism for isolated impurities, and we find that even when disorder takes on a more complex form, this correlation between the order parameter and the coherence-peak height persists. We find that this mechanism describes these particular correlations in FeTeSe well, but we find that for BSCCO this picture needs to be bolstered by strong-coupling effects to account fully for the material's STS phenomenology, in particular the necessity of a spatially dependent scattering rate (presumably due to interaction effects) that is neglected in our disorder-only model.\cite{alldredge2008evolution}

\section{Model and Methods} \label{model}

In this section, we will discuss the model and methods used in the study of the correlations between the superconducting order parameter and various spectroscopic quantities that can be extracted from STS experiments. Our starting point is a square-lattice tight-binding model with attractive nearest-neighbor interactions. The Hamiltonian is
\begin{equation}
H = -\sum_{ij\sigma} t_{ij}c_{i\sigma}^{\dagger}c_{j\sigma} + \frac{V_0}{2}\sum_{\langle ij \rangle \sigma\sigma'}c_{i\sigma}^{\dagger}c_{i\sigma}c_{j\sigma'}^{\dagger}c_{j\sigma'}.
\label{eq:hubbardhamiltonian}
\end{equation}
$\langle ij \rangle$ in the second term of Eq.~\ref{eq:hubbardhamiltonian} signifies that the sum over $i$ and $j$ is restricted to nearest-neighbor pairs of sites. Treating interactions within mean-field theory, we define $\Delta(i, j) = V_0\langle c_{i\uparrow} c_{j\downarrow}  \rangle$, where $i$ and $j$ are nearest-neighbor sites;  this leads us to the following mean-field Hamiltonian describing a $d$-wave superconductor:
\begin{equation}
H = -\sum_{ij\sigma} t_{ij}c_{i\sigma}^{\dagger}c_{j\sigma} + \sum_{ij}\large(\Delta(i,j)^{\ast}c_{i \uparrow}c_{j \downarrow} + \text{h.c.}\large).
\label{eq:hamiltonian}
\end{equation}
The hopping matrix elements are
\begin{equation}
    t_{ij} = \left \{ \begin{array}{ll} 
    V_\mathrm{imp}(i), & \mbox{$i=j$} \\
    t, & \mbox{$i$ and $j$ are n.n.} \\
    t', & \mbox{$i$ and $j$ are n.n.n.} \end{array} \right .,
\end{equation}
where $V_{imp}(i)$ is the impurity potential on site $i$, $t = 1$ and $t' = -0.3$ are the nearest-neighbor (n.n.) and next-nearest-neighbor (n.n.n.) hopping matrix elements, respectively.  Throughout this work, we choose the chemical potential so that the hole doping of the clean system is $10\%$, relative to half-filling.  Because the chemical potential, rather than the electron density, is fixed, the impurity potential will dope the system; however, in all cases we choose $V_\mathrm{imp}(i)$ such that the doping is small.

To obtain the LDOS and the superconducting order parameter efficiently for large system sizes, we use a Green's function formalism. The Green's function $G(i,j, \omega)$ is defined as 
\begin{equation}
G(i,j, \omega) = [\omega + i\eta - H]^{-1}_{i,j},
\label{eq:greensfunction}
\end{equation}
where $[]^{-1}$ is a matrix inverse in Nambu space, and $\eta$ is a small broadening parameter that we take to be a constant. $G$ and $H$ are both $2N_xN_y \times 2N_xN_y$ matrices, with $N_x$ and $N_y$  the dimensions of the system in the $x$- and $y$-directions, respectively. By imposing periodic boundary conditions along the $y$-direction and open boundary conditions along the $x$-direction, one can rewrite $H$ to be block diagonal. This implies that the diagonal subblocks of $G$, and hence the LDOS, can be obtained very efficiently using a recursive algorithm described in detail elsewhere in the literature.\cite{godfrin1991method, reuter2012efficient, sulangi2017revisiting, sulangi2018quasiparticle} Consequently, very large system sizes with $\mathcal{O}(10^5)$ sites are accessible with this method.

The order parameter is obtained from the self-consistent solution of
\begin{equation}
\Delta(i,j) = -\frac{V_{ij} }{\pi}\int_{-\infty}^{\infty}d\omega \, n_f(\omega, T) \text{Im}\left[G_{12}(i,j, \omega)\right],
\label{eq:orderparam}
\end{equation}
 where
 \begin{equation}
     V_{ij} = \left \{ \begin{array}{ll} V_0, & i,j \mbox{ are  n.n.} \\
     0, & \mbox{otherwise} \end{array} \right .,
 \end{equation}
 is the pairing interaction, and $n_f(\omega, T)$ is the Fermi function. Eq.~\ref{eq:orderparam} appears to require the evaluation of the off-diagonal  (in real space) blocks of $G$. However, since we are interested only in nearest-neighbor $d$-wave pairing, we merely need to obtain the elements of $G$ corresponding to nearest-neighbor pairs, which requires the evaluation of only the lower and upper diagonal subblocks of $G$, and hence the amount of computational time does not scale up dramatically; these subblocks can be obtained by recursion from the diagonal subblocks which are the first set of outputs of the algorithm. The main obstacle turns out to be the frequency integral, which requires a wide range of energies over which $G$ is calculated; however, the calculational effort scales only linearly in the number of frequencies used and is therefore manageable under most circumstances. All our calculations are performed in the limit $T \rightarrow 0$; we focus on this limit because the experimental JSTS studies on unconventional superconductors published thus far have been performed at very low temperatures deep within the superconducting state.\cite{hamidian2016detection,cho2019strongly}  The possibility that the pairing interaction itself is disordered was discussed previously;\cite{nunner2005dopant,nunner2006fourier,Roemer2018} we neglect this possibility here, but note that $\Delta(i,j)$ is disordered in response to the random impurity potentials discussed in this paper. 

We are primarily interested in four main quantities: the $d$-wave superconducting order parameter; the Josephson critical current $I_c$ obtained from Cooper-pair tunneling from a $d$-wave superconducting tip to a $d$-wave superconducting sample; the spectral gap obtained from differential conductance measurements; and the height of the coherence peaks, also obtained from differential conductance measurements. The $d$-wave component of the order parameter is computed on each lattice site as
\begin{equation}
    \Delta_d(i) = \sum_{\boldsymbol{\delta}} (-1)^{\delta_y} \Delta(i,i+\boldsymbol{\delta}),    
\end{equation}
where $\boldsymbol{\delta} = (\delta_x,\delta_y)$ connects $i$ to its four nearest neighbors. For the calculation of $I_c$, we follow Graham and Morr,\cite{graham2019josephson} with
\begin{equation}
 I_c = 2\frac{4e}{\hbar}t^2_0 \int_{-\infty}^{\infty}\frac{d\omega}{2\pi} n_f(\omega, T) W( \omega).
 \label{eq:criticalcurrentspin}
\end{equation}
Here, $t_0$ is the tunneling amplitude between the tip and the sample and $W(\omega)$ is 
\begin{eqnarray}
  W(\omega) = \sum_{i,j}\text{Im} \large[ G_{12}^{tip}(i,j, \omega) \nonumber  G_{12}^{sample}(j,i, \omega)\large].
  \label{eq:criticalcurrentkernel}
\end{eqnarray}
 To simplify matters, we have taken the tip to be a site-centered ``filter'' with five atoms in the shape of a cross (this is the smallest tip one can make which measures $d$-wave correlations in an $x$-$y$-symmetric fashion). We further assume this tip to be made of the same $d$-wave superconductor, but \emph{without} disorder, as was done in Ref.~\onlinecite{graham2019josephson}. 
  
\begin{figure}[t]
	\centering
	\includegraphics[width=0.5\textwidth]{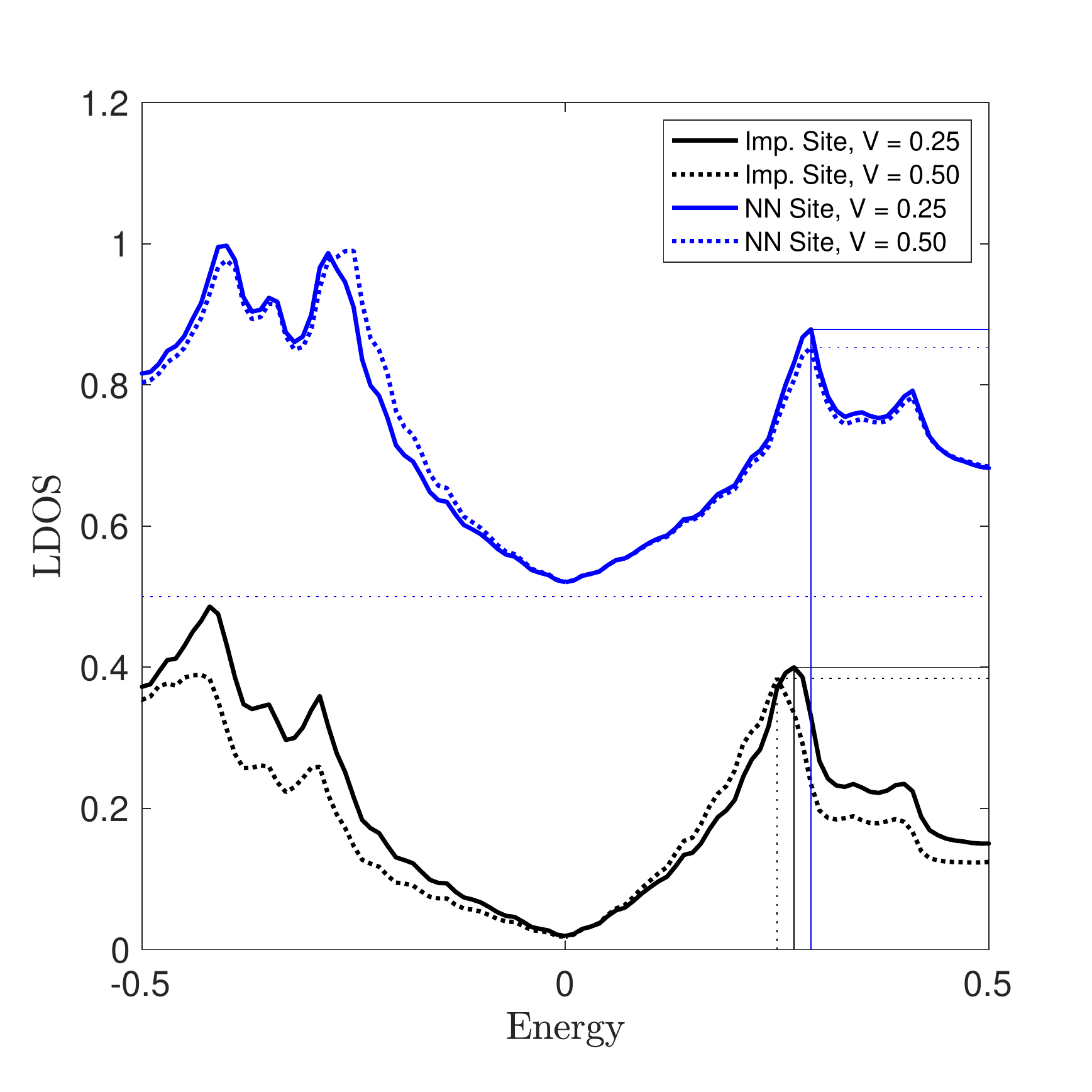}\\
	\caption{Plots of the LDOS vs. energy on two sites (impurity site, black, and nearest-neighbor site, blue) for a single impurity with strength $V = 0.25$ (solid line) and $V = 0.50$ (dashed line), showing how the quantities shown in Fig.~\ref{fig:singleimpuritycombinedplot} are obtained. Vertical lines show the spectral gap, while horizontal lines show the height of the coherence peaks. The blue horizontal dashed line denotes the baseline for the nearest-neighbor LDOS plots. Energies are expressed in units of $t$, and LDOS in states/$t$/unit cell.}
	\label{fig:ldosnearimpurity}
\end{figure}

For each position $i$, we obtain the spectral gap and coherence peak height from LDOS at $i$.  This is illustrated in Fig.~\ref{fig:ldosnearimpurity}, which shows LDOS spectra for sites on and adjacent to an isolated weak-scattering impurity.  As shown in the figure, the coherence peak height is given by the largest value of the LDOS at positive energies, and the spectral gap is the energy at which the peak occurs.  We focus on the LDOS at positive energies to avoid complications arising from a van Hove singularity that is found at negative energies.  Note that the shifts in peak height and spectral gap in Fig.~\ref{fig:ldosnearimpurity} are relatively small but, as we show below, are substantially larger when the impurity density is high.

Throughout this paper we are interested in the correlations among scanning tunneling spectroscopy (STM) observables. To quantify this, we use the correlation coefficient $r$, which is defined for two real-space quantities $P_i$ and $Q_i$ with similar dimension as
  \begin{equation}
  r = \frac{\sum_{i}(P_{i} - \bar{P})(Q_{i} - \bar{Q})}{\sqrt{\sum_{i}(P_{i} - \bar{P})^2} \sqrt{\sum_{i}(Q_{i} - \bar{Q})^2)}},
  \label{eq:correlationcoefficient}
  \end{equation}
  where $\bar{P}$ and $\bar{Q}$ are the spatial averages of $P_i$ and $Q_i$ and the sums run over spatial sites. This is the same definition used by Cho \emph{et al.} in their analysis of JSTS data.\cite{cho2019strongly} In our calculations, we use a system consisting of $1000\times50 = 50000$ sites, and use the middlemost $336\times48$ subsection of the system for the calculation of $r$. This choice gives us over $16000$ distinct values of $P_i$ and $Q_i$ in Eq.~\ref{eq:correlationcoefficient}, which is large enough for our correlation analyses to be statistically meaningful.
 
\section{Pointlike Impurities}
\label{sec:pointlike}

\begin{figure}[t]
	\centering
	\includegraphics[width=0.5\textwidth]{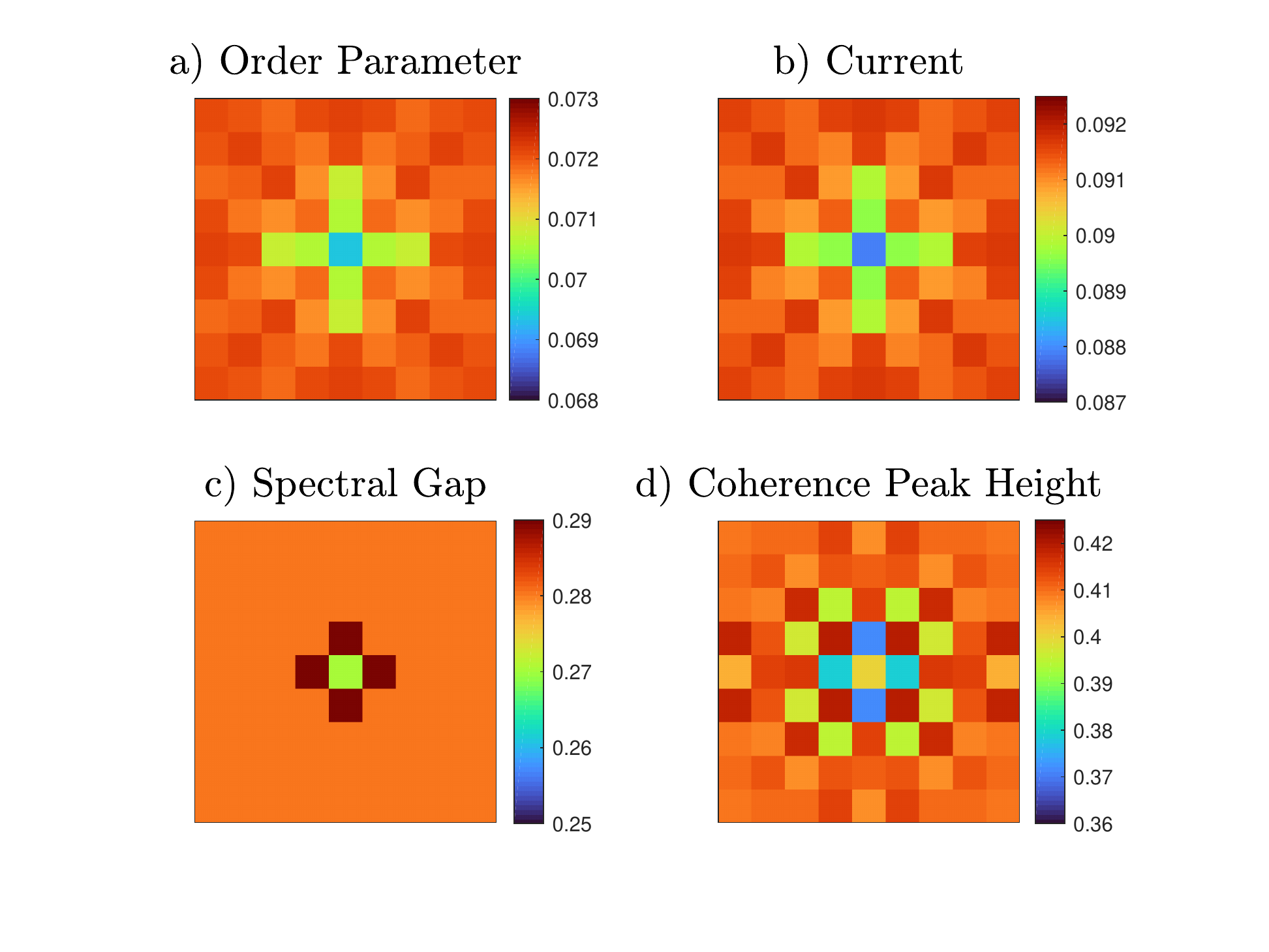}\\
	\caption{Plots of a) the $d$-wave order parameter, b) $I_c$ (in units of $4et^2_0/\hbar$), c) the spectral gap, and d) the coherence-peak height taken within a region with a single isolated impurity. It can be seen that $I_c$ is almost identical to the true order parameter, while the spectral gap hardly resembles the true order parameter.}
		\label{fig:singleimpuritycombinedplot}
\end{figure}

In this section, we explore the correlations between the four quantities of interest---the order parameter, the Josephson current $I_c$, the spectral gap, and the coherence peak height---that are induced by a concentration $p$ of pointlike impurities.   The impurity potential takes the form 
\begin{equation}
    V_\mathrm{imp}(i) = \left \{ \begin{array}{ll} 
    V, & i \in \{i\}_\mathrm{imp}, \\
    0, & \mbox{otherwise},
    \end{array} \right .,
\end{equation}
where $\{i\}_\mathrm{imp}$ is the set of lattice sites that host an impurity. Two limits,  the low  and high impurity concentration limits, can be understood qualitatively, and these are discussed in turn below. This form of disorder is very well-studied and can be understood analytically in the weak-impurity Born limit and the unitary
limit of the disorder-averaged theory,\cite{gorkov1985defects,hirschfeld1988consequences,lee1993localized,durst2000impurity} but it is also known that nontrivial multi-impurity effects not amenable to analytical treatment naturally result when the concentration of impurities becomes sufficiently large.\cite{zhu2003two,atkinson2003quantum}

In the dilute limit, the correlations are determined by the spatial patterns of the quantities of interest around each impurity.  We thus start with a discussion of a single weak-scattering impurity in isolation. We take the impurity potential to be $V=0.25$ and calculate the superconducting order parameter self-consistently. In Fig.~\ref{fig:singleimpuritycombinedplot}a, it can be seen  that the order parameter is reduced slightly at the impurity site and relaxes towards its clean-limit value over a length scale of a few lattice spacings. The corresponding Josephson current (Fig.~\ref{fig:singleimpuritycombinedplot}b) has an almost identical spatial pattern to the order parameter, which confirms a similar result obtained by Graham and Morr.\cite{graham2019josephson} This is not only the case for isolated weak-scattering impurities; rather, the cross-correlation coefficient $r$ between the order parameter and $I_c$ is consistently very high ($r \approx 0.99$) for all disorder types and strengths we have studied. $I_c$ is thus an almost perfect indicator of the spatial dependence of the order parameter. In the Appendix we provide evidence for the near-perfect matching between the order parameter and $I_c$ across various disorder types and strengths.

In contrast, both the spectral gap (Fig.~\ref{fig:singleimpuritycombinedplot}c) and coherence peak heights (Fig.~\ref{fig:singleimpuritycombinedplot}d) have spatial patterns that differ visibly from that of the order parameter.  The influence of the impurity potential on the spectral gap is short-ranged, extending only to the adjacent site where the spectral gap is enhanced relative to its bulk value.  Since the order parameter is reduced on sites adjacent to the impurity, there is a weak negative correlation between the order parameter and the spectral gap.  However, as shown below, this result is not universal and depends on the details of the impurity potential and the amount of disorder.  

The coherence-peak height (Fig.~\ref{fig:singleimpuritycombinedplot}d) has a relatively complex spatial pattern. Like the  order parameter, it is reduced at the impurity site and relaxes to its bulk value within a few lattice sites. Unlike the order parameter, there are additional short-wavelength oscillations of the peak height. Despite this difference, the coherence peak height has a strong positive correlation with the order parameter and a corresponding strong negative correlation with the SG.  As we show below, these correlations persist up to large impurity concentrations.

Next, we consider an ensemble of weak pointlike impurities that are randomly distributed throughout the sample, such that each lattice site has a probability $p$ of hosting an impurity with on-site potential strength $V=0.25$. Since we are at fixed chemical potential, the impurities change the electron density.  At the largest impurity concentration considered in this section, $p=20\%$, we estimate that this dopes the system by $\sim 0.01$ electrons per unit cell, which is negligible.

\begin{figure*}[tb]
	\centering
	\includegraphics[width=1.0\textwidth]{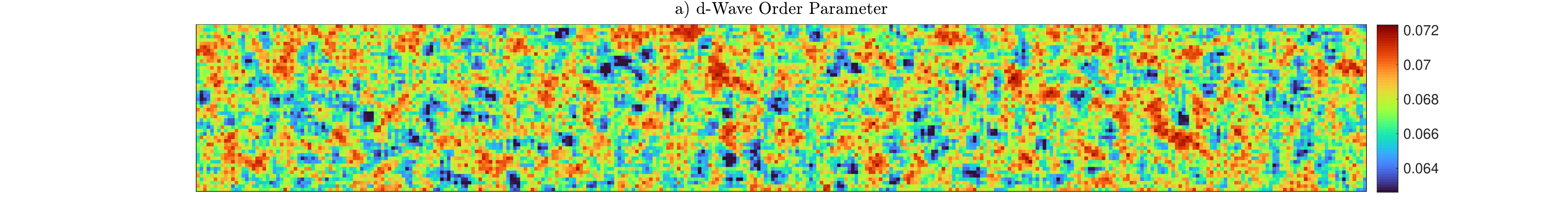} \\
	\includegraphics[width=1.0\textwidth]{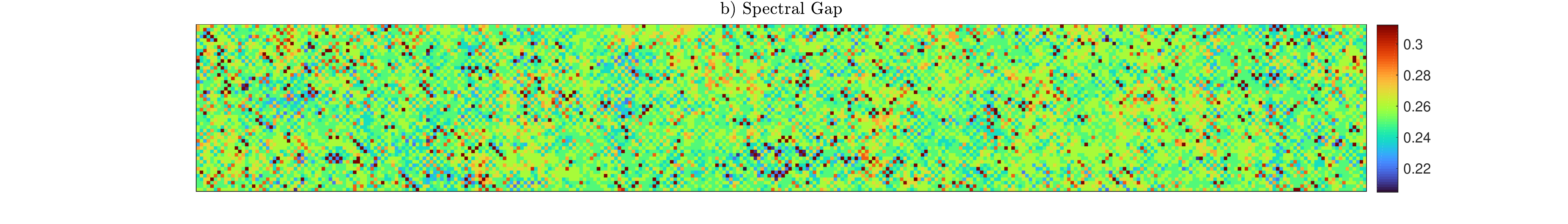}  \\
	\includegraphics[width=1.0\textwidth]{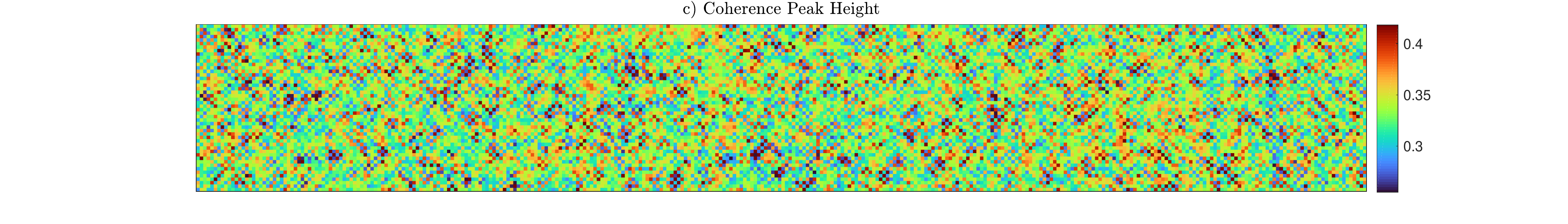}  
	\caption{Plots of (top to bottom) the $d$-wave order parameter, the spectral gap, and the coherence-peak height for a $d$-wave superconductor with weak pointlike impurities with strength $V = 0.25$ and concentration $p = 20\%$.}
	\label{fig:pointlikecombinedplot}
\end{figure*}

Figure~\ref{fig:pointlikecombinedplot} shows the spatial patterns of the $d$-wave order parameter, the spectral gap, and the coherence peak height, for a high concentration ($p=20\%$) of weak-scattering impurities.  Such an impurity distribution might, for example, be a model for Sr ions in overdoped La$_{2-x}$Sr$_x$CuO$_4$.\cite{Eisaki2005,lee2020low, li2021superconductor}
Certain similarities with the single impurity case are evident in the figure:  Notably, the order parameter has a smooth spatial profile, while the spectral gap responds to the impurity potential on a short length scale and the coherence peak height shows short-range oscillations on top of a smooth envelope. There are also important differences: notably, the variations of both the spectral gap and the coherence peak height are significantly larger here than for the single-impurity case, although the range of order parameter values is about the same.

\begin{figure}[tb]
	\centering
	\includegraphics[width=0.5\textwidth]{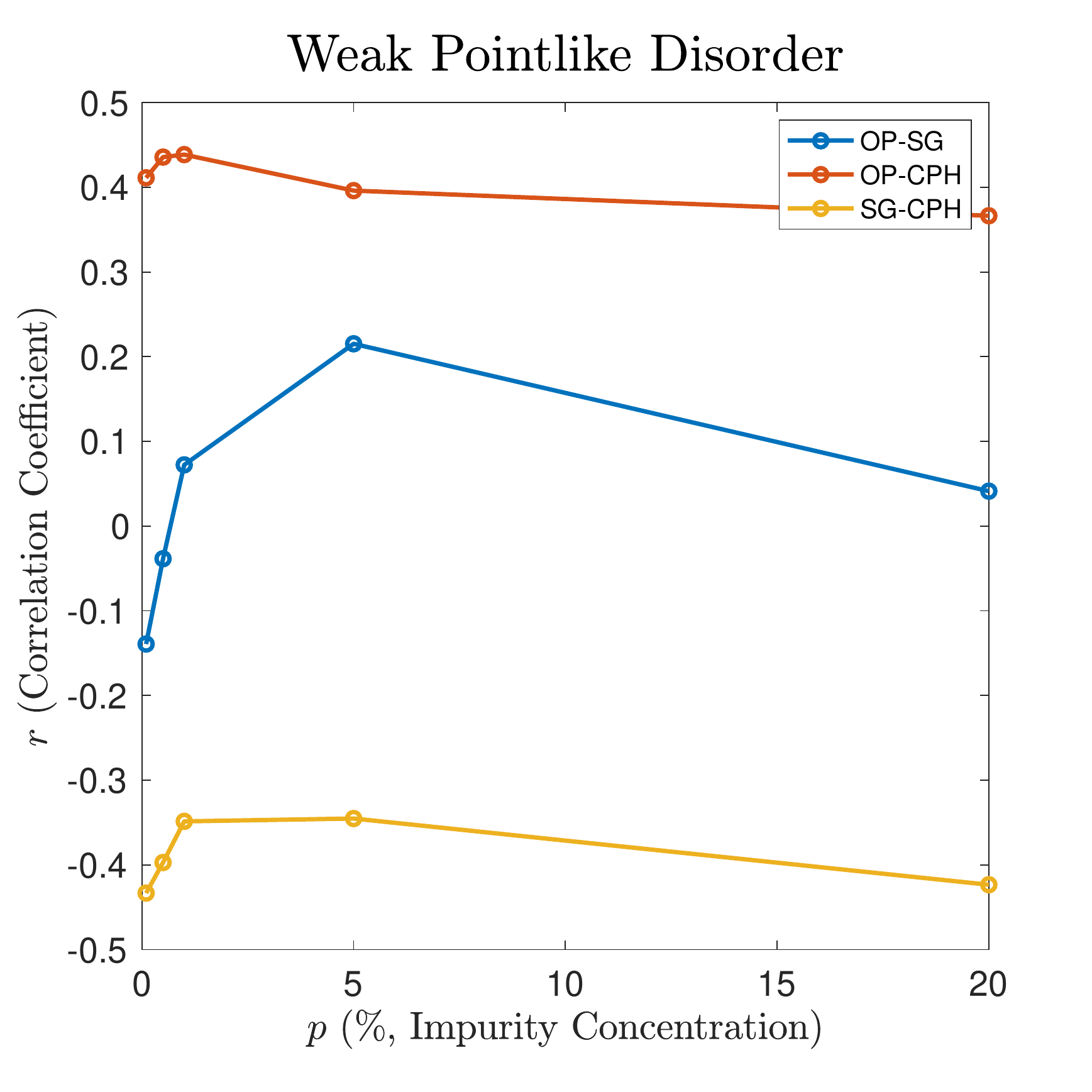}\\
	\caption{Correlation coefficients $r$ for various pairs of quantities as a function of impurity concentration $p$. Here the impurities are, as described in the main text, weak scatterers with $V = 0.25$.}
	\label{fig:pointcorr}
\end{figure}

At low impurity concentrations, the similarities with the single-impurity case are reflected in the correlations between the different quantities of interest (Fig.~\ref{fig:pointcorr}): The correlation coefficient $r$ between the order parameter and the spectral gap (OP-SG) is small and negative;  the correlations between the order parameter and the coherence-peak height (OP-CPH) are large and positive; and the correlations between the spectral gap and coherence peak height (SG-CPH) are large and negative.  For each of these,  $r$ measures correlations between individual patterns near isolated impurities, as in the single-impurity case.  

Two of the correlation functions (SG-CPH and OP-CPH) are nearly independent of impurity concentration, while the third (OP-SG) depends significantly on $p$, changing from negative to positive and then decreasing towards zero as $p$ increases.  The small value of the OP-SG correlation coefficient at high impurity concentrations can be explained by the differing length scales over which the order parameter and spectral gap respond to the impurities: at a position $i$, the order parameter depends on the distribution of impurities within a coherence length of $i$, while the spectral gap responds locally to individual impurities.   The two quantities are thus uncorrelated when the number of impurities in a correlation volume is large {(note that in our simulations, the average BCS coherence length of the clean system is about $\xi_0\simeq 3$ lattice constants).  }

In contrast, correlations between the coherence peak height and the order parameter (OP-CPH) or spectral gap (SG-CPH) are both large and nearly independent of doping. 
It appears that the intuitive correlations one can obtain from Fig.~\ref{fig:singleimpuritycombinedplot}c and d continue to hold even in a multi-impurity setting: the correlations involving CPH involve mainly the nearest-neighbor sites, where both the CPH and the OP are suppressed but the SG is enhanced. Since the main effect of the CPH on the cross-correlations is localized on a small number of sites surrounding each impurity, these correlations are largely independent of $p$.

\section{Binary-Alloy Disorder} 
\label{sec:binaryalloy}

\begin{figure*}[ht]
	\centering
	\includegraphics[width=1.0\textwidth]{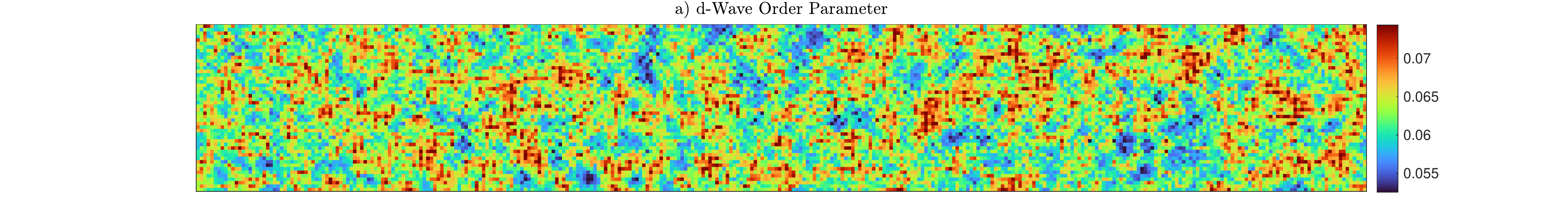} \\
	\includegraphics[width=1.0\textwidth]{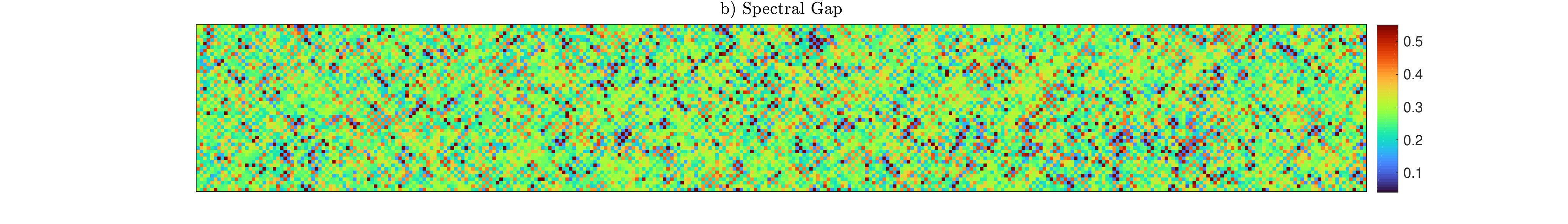}  \\
	\includegraphics[width=1.0\textwidth]{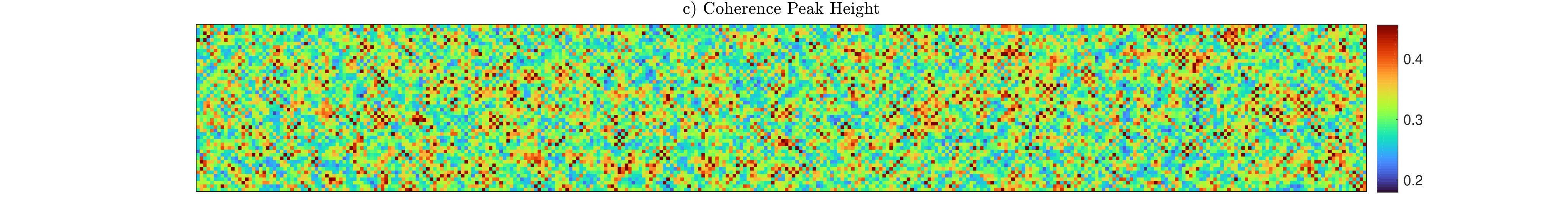}  
	\caption{Plots of (top to bottom) the $d$-wave order parameter, the spectral gap, and the coherence-peak height for a $d$-wave superconductor with binary-alloy disorder with strength $V_b = 0.250$.}
	\label{fig:binaryalloycombinedplot}
\end{figure*}

\begin{figure*}[tb]
	\centering
	\includegraphics[width=0.15\textwidth]{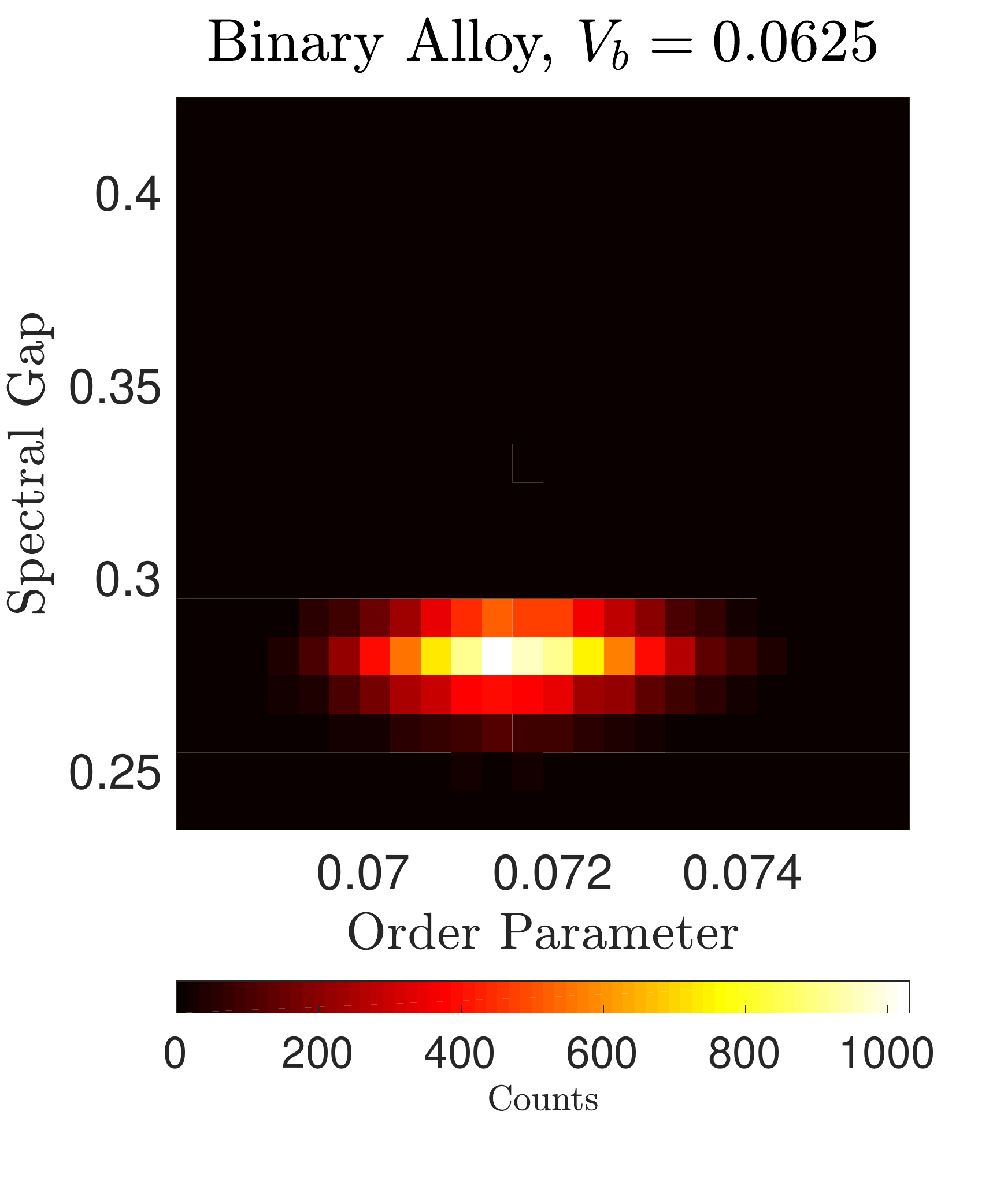}\quad
	\includegraphics[width=0.15\textwidth]{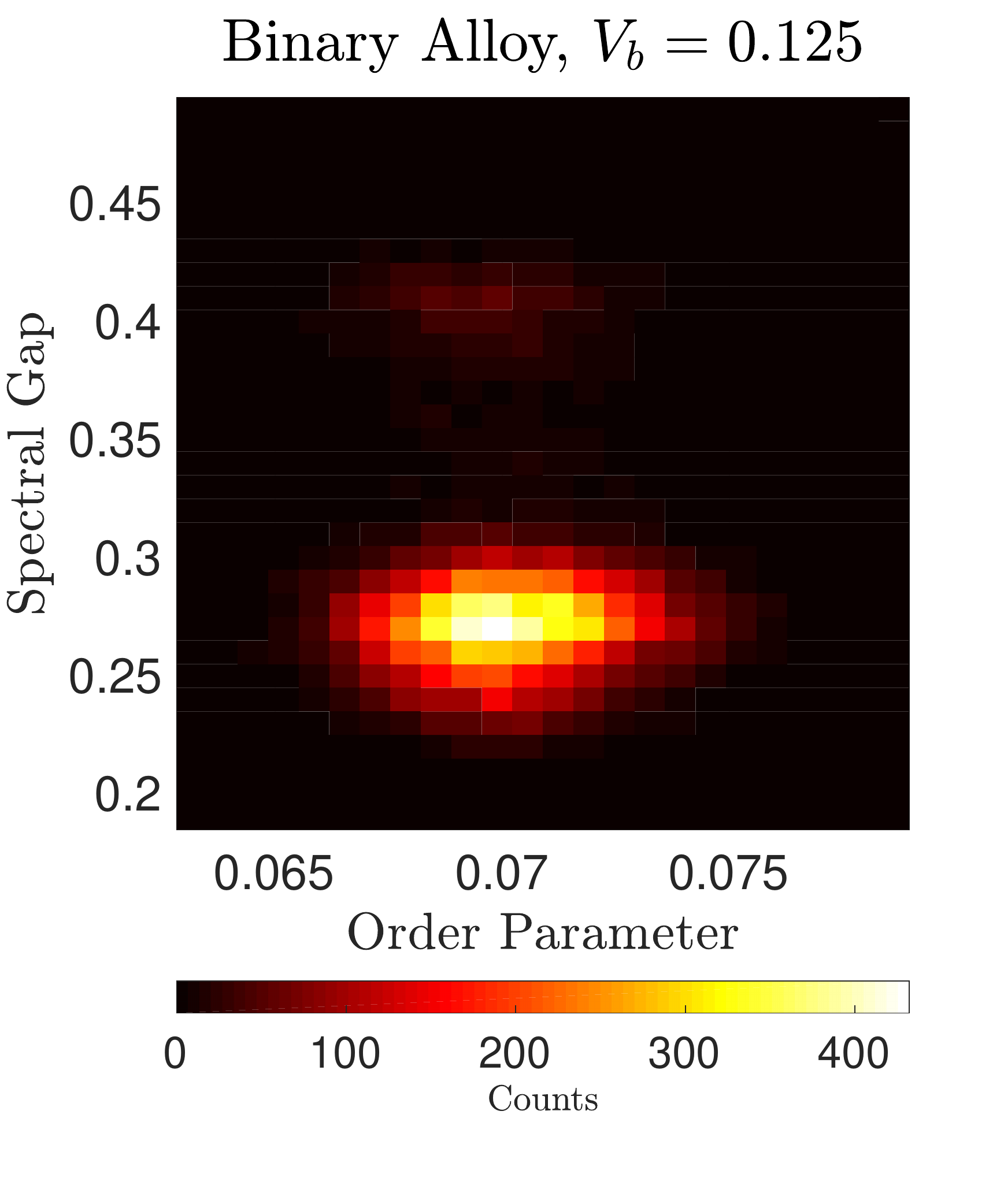}\quad
	\includegraphics[width=0.15\textwidth]{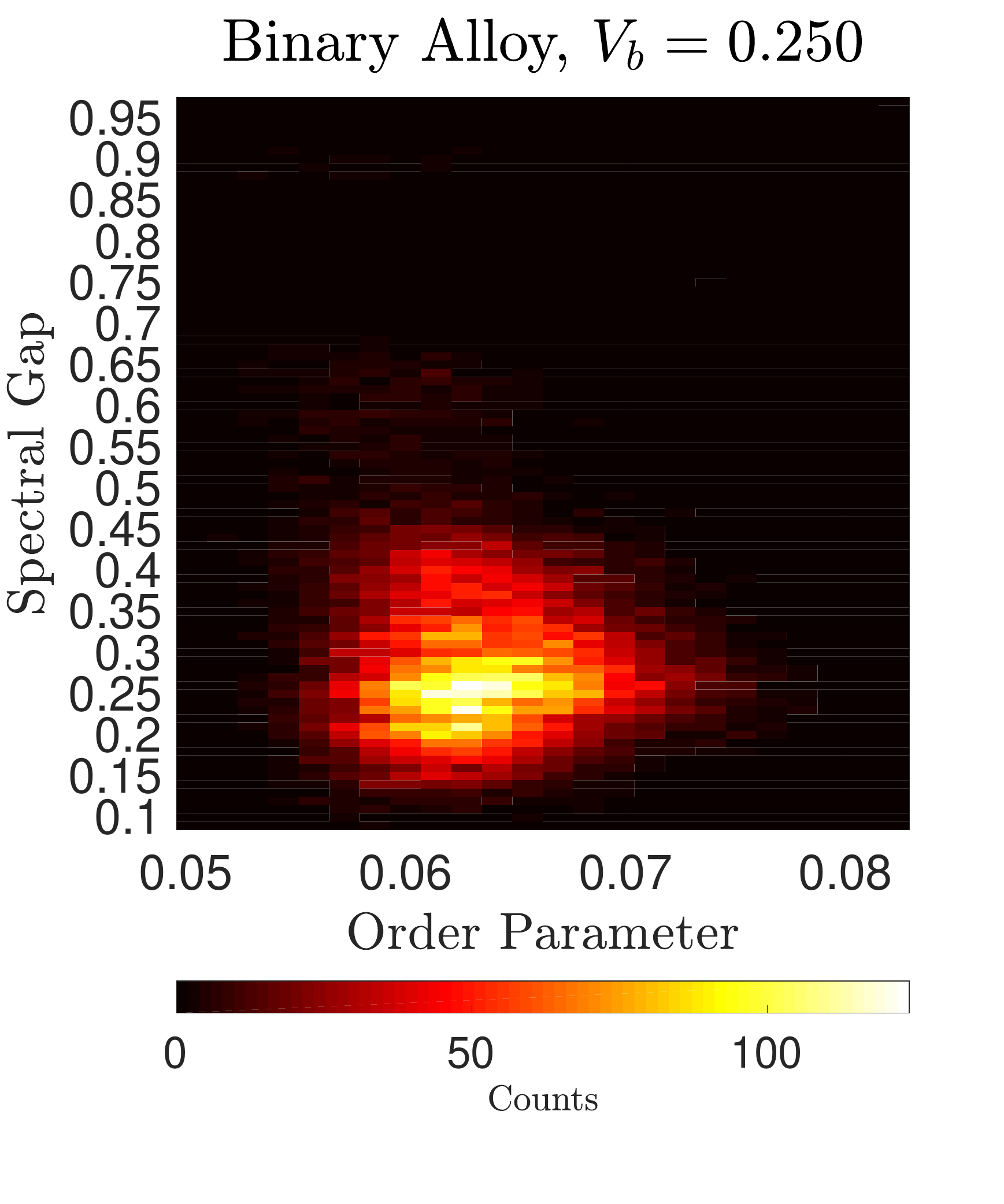}\quad
	\includegraphics[width=0.15\textwidth]{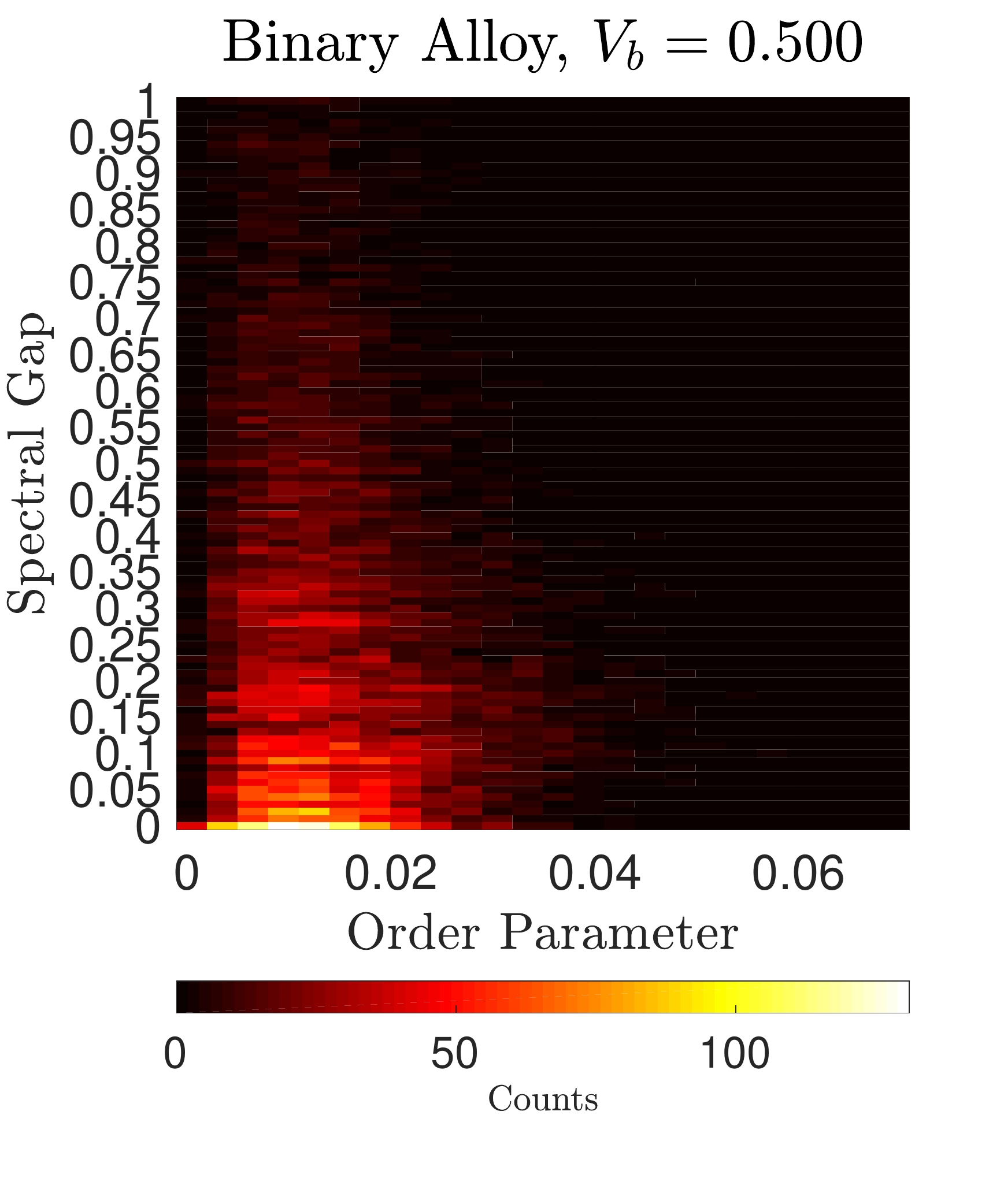}\\
	\includegraphics[width=0.15\textwidth]{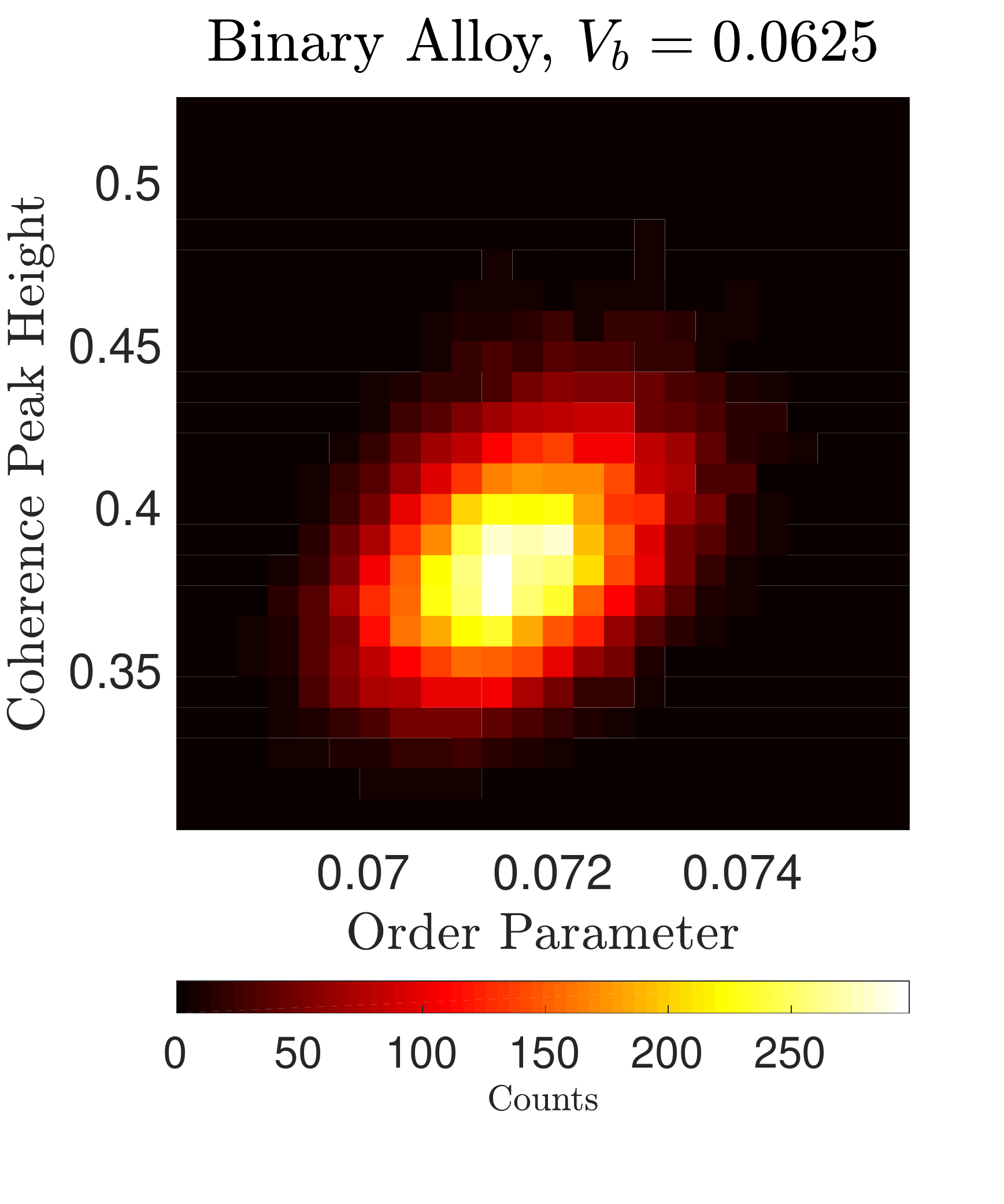}\quad
	\includegraphics[width=0.15\textwidth]{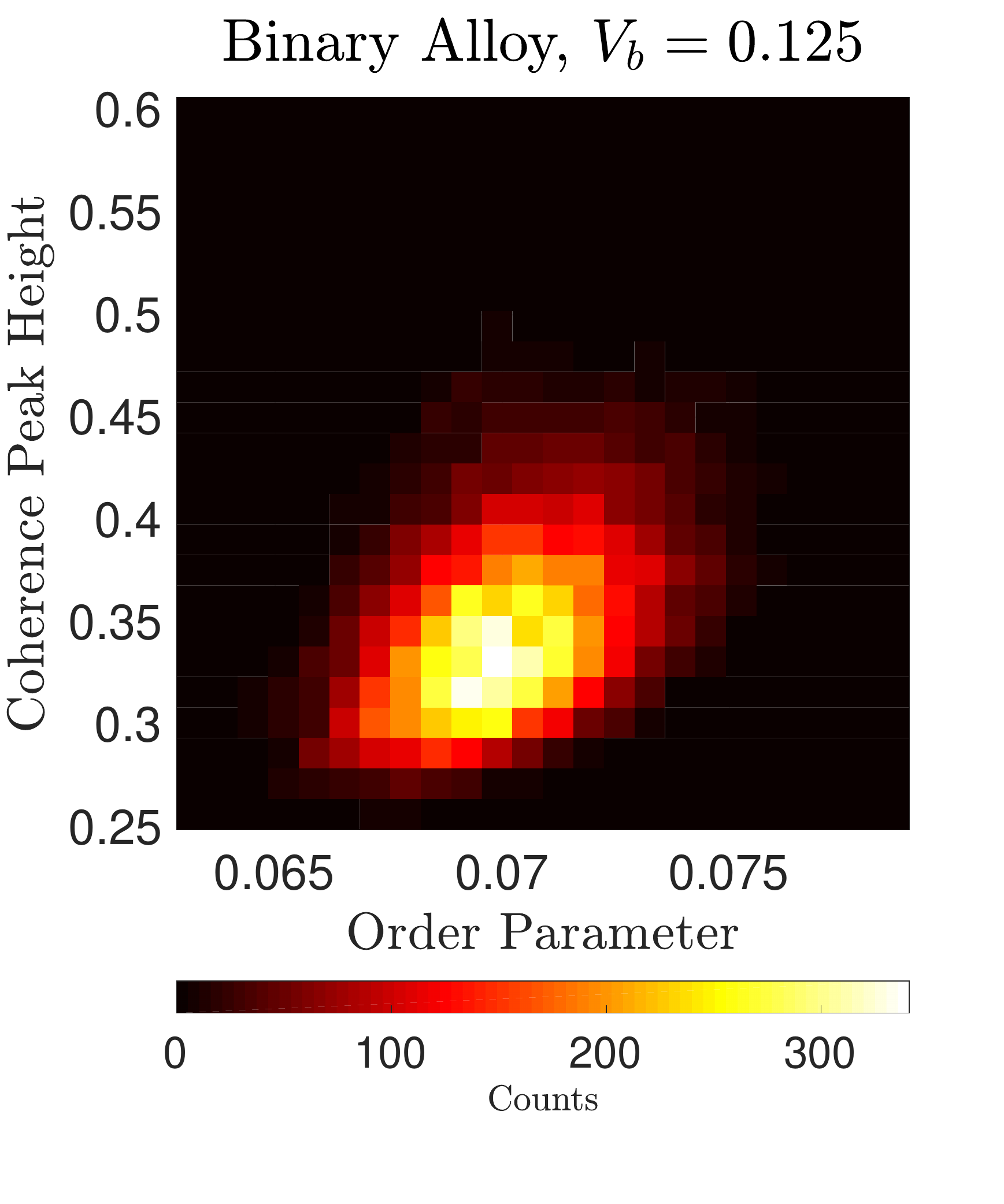}\quad
	\includegraphics[width=0.15\textwidth]{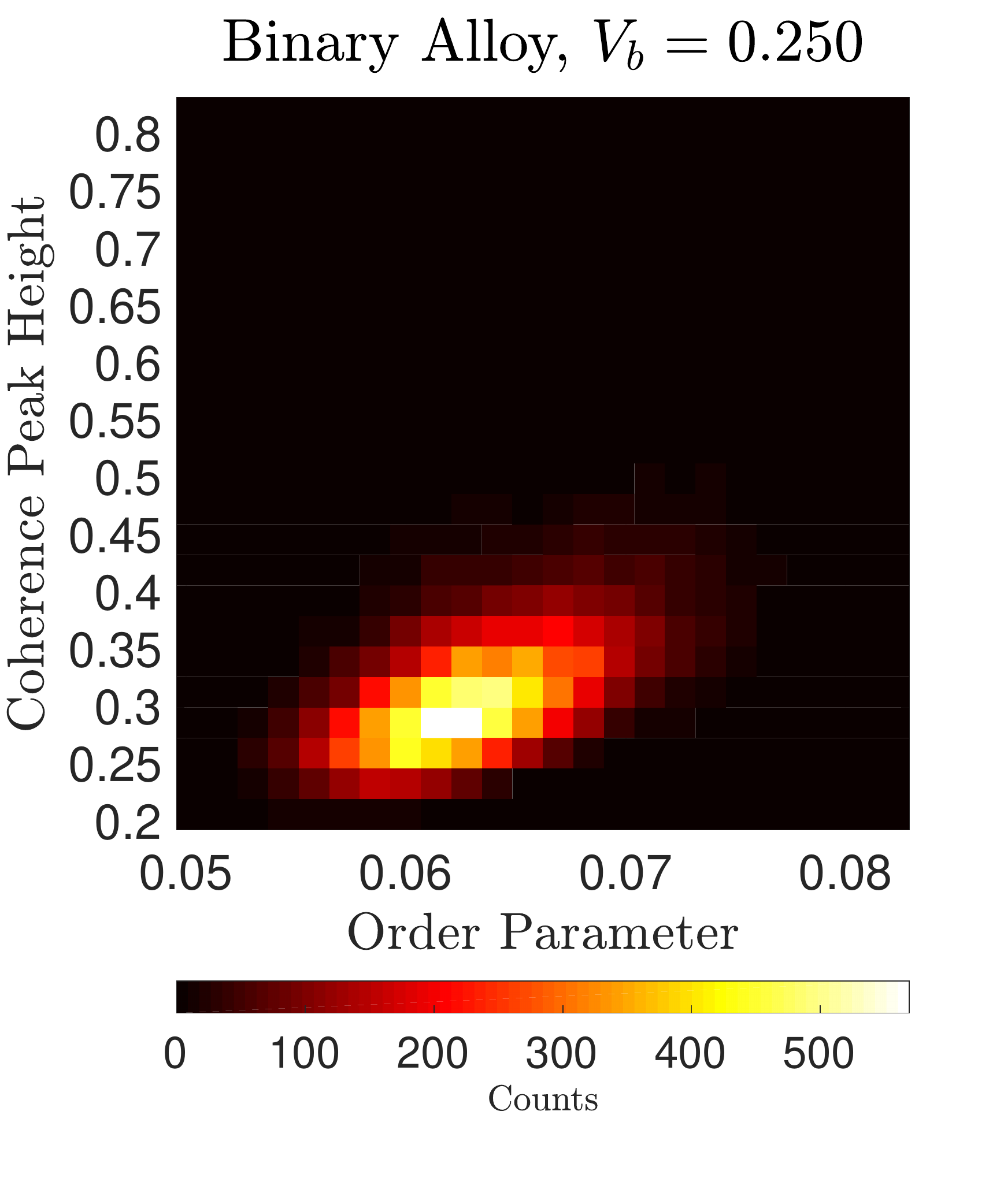}\quad
	\includegraphics[width=0.15\textwidth]{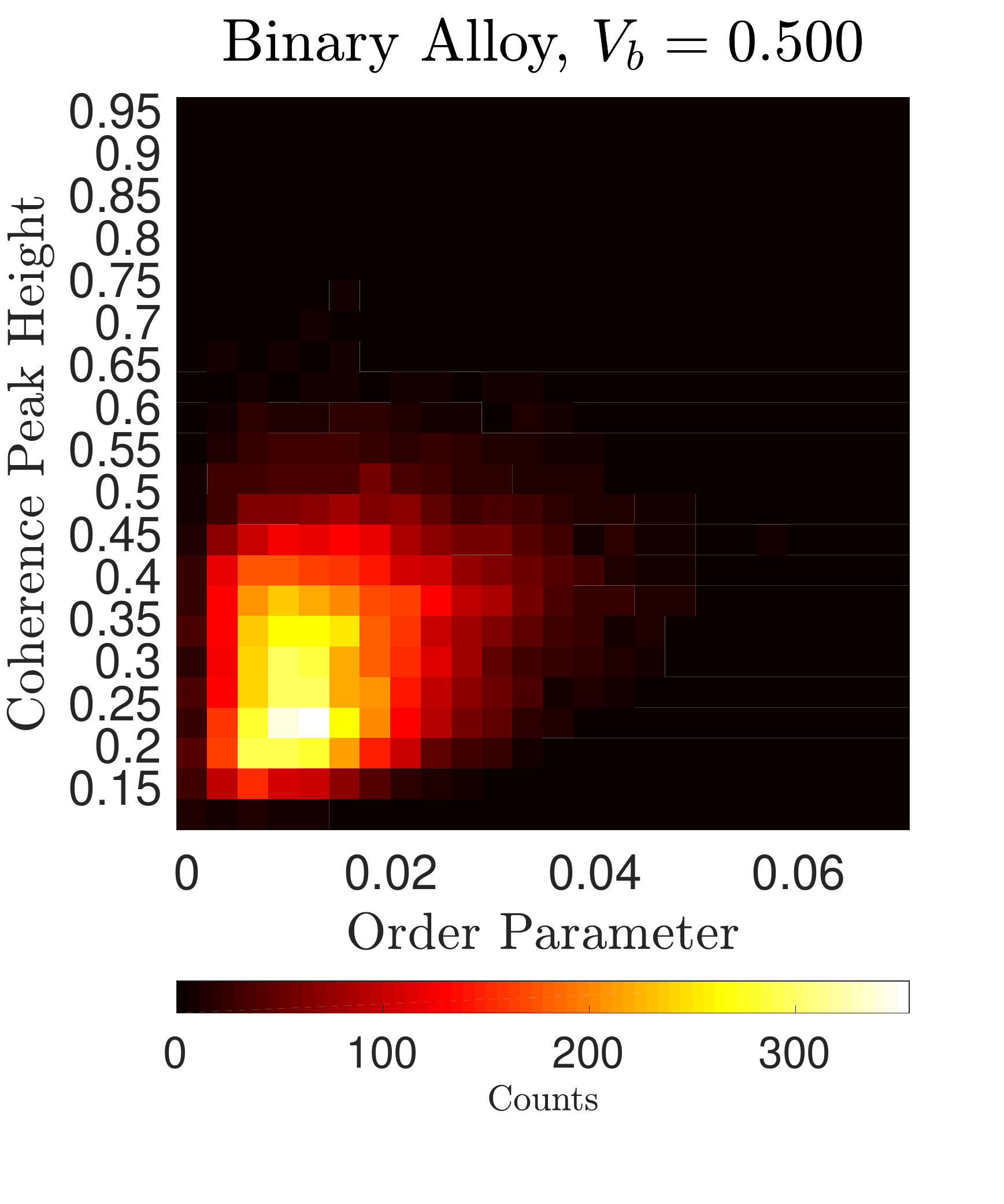}\quad
	\caption{Two-dimensional histograms between the order parameter and the spectral gap (top row) and the order parameter and the coherence-peak height (bottom row), shown for varying binary-alloy disorder strength $V_b$ (left to right). Note that the scales of the $x$- and $y$-axes are not the same as $V_b$ increases.}
	\label{fig:binaryalloyhistograms}
\end{figure*}

\begin{figure}[bt]
	\centering
	\includegraphics[width=0.5\textwidth]{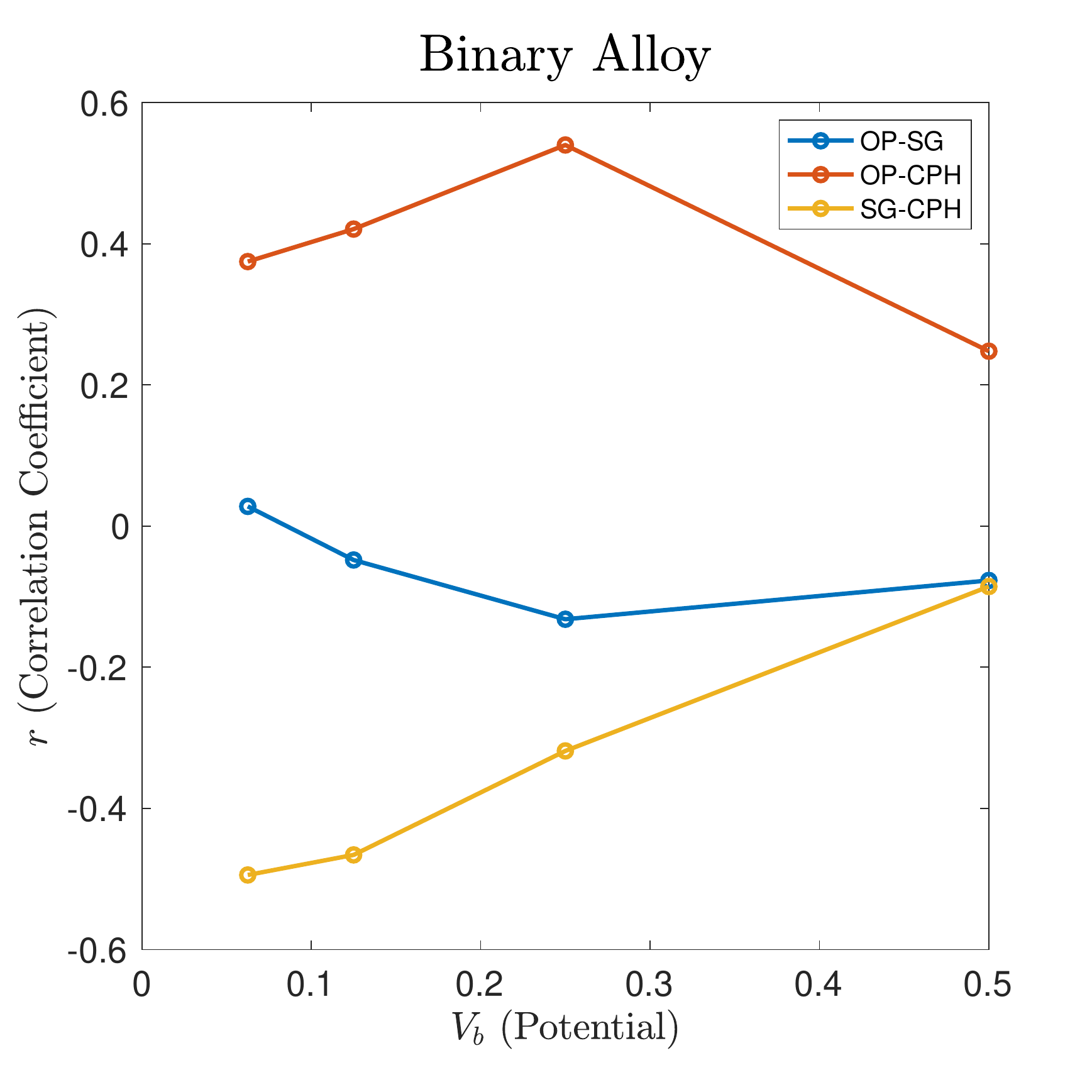}\\
	\caption{Correlation coefficients $r$ for various pairs of quantities as a function of binary-alloy impurity strength $V_b$.}
	\label{fig:binaryalloycorr}
\end{figure}

We next consider a binary-alloy model, in which each lattice site has an equal probability of hosting one of two ions, with ionic potentials $\pm V_b$.  Such a model might be appropriate for the iron-based superconductor FeTeSe, in which the Te and Se ions form a solid solution, and which is a highly inhomogeneous superconductor with no easily identified correlations between topographic and electronic maps.\cite{cho2019strongly} This disorder model had previously been employed by Berthod in a numerical study of vortices in FeTeSe.\cite{berthod2018signatures} To the best of our knowledge, no analytical studies on superconductivity employing this model have been performed. This form of disorder can still be understood within the weak-impurity Born limit as long as $V_b$ is very small. However, for stronger impurities, no such analytical treatment exists, since the concentration of impurities in this case is so large that the standard $T$-matrix approximation for multi-impurity systems ceases to be valid.

While we keep the relative proportions of each ionic component fixed, we tune $V_b$ from 0.0625 to 0.5, which covers the evolution of disorder from the weak Born limit to the strongly disordered limit where superconductivity is strongly suppressed. We use the same tight-binding parameters here as in the previous simulations and keep the background chemical potential fixed. While our model consists of a single band with $d$-wave pairing, we believe that the qualitative aspects should carry over to systems with a general sign-changing gap order parameter, like $s_\pm$ in Fe-based, multiband superconductors such as FeSeTe.

In Fig.~\ref{fig:binaryalloycombinedplot} we show spatially resolved plots of the $d$-wave order parameter, the spectral gap, and the coherence-peak height for $V_b = 0.25$.  Similar to what was seen in Fig.~\ref{fig:pointlikecombinedplot}, each quantity has its own characteristic response to the impurity potential. The order parameter has a patchy structure that emerges despite the sharply varying atomic-scale disorder present in the system; the spectral gap has pronounced variations on the atomic length scale; and the coherence peak height exhibits short wavelength oscillations on top of a smoothly varying envelope.  However, because the density of impurities is high, there is no limit in which one can make a direct connection to the single-impurity case.  Indeed, as we show below, there are important differences with the pointlike impurity model discussed in Sec.~\ref{sec:pointlike}.

Two-dimensional histograms showing both the $d$-wave order parameter and either the spectral gap or the coherence-peak height are shown in Fig.~\ref{fig:binaryalloyhistograms} for several different values of $V_b$. The key result for this figure is that there is a clear positive correlation between the order parameter and the coherence peak height, while the spectral gap is, at best, weakly correlated with the order parameter.  This is similar to what was found for pointlike impurities; however, the dependence on the amount of disorder is different.

This is illustrated by Fig.~\ref{fig:binaryalloycorr}, which shows the OP-SG, OP-CPH, and SG-CPH correlation coefficients as a function of the impurity potential.  The OP-CPH correlation coefficient is near $r=0.4$ at small $V_b$, and grows with increasing $V_b$, except for the last point at $V_b=0.5$.  This is consistent with the obvious increase of the correlation between the two quantities in Fig.~\ref{fig:binaryalloyhistograms} with increasing $V_b$.   Figure~\ref{fig:binaryalloyhistograms} also reveals that the drop in correlations at $V_b=0.5$ is connected to the suppression of superconductivity by the large disorder potential.  

Figure~\ref{fig:binaryalloycorr} also reveals that the OP-SG correlation coefficient is small, which is again consistent with the absence of any obvious correlation in Fig.~\ref{fig:binaryalloyhistograms}.  Finally, Fig.~\ref{fig:binaryalloycorr} shows that the SG-CPH correlation coefficient is negative and large at weak disorder, but decreases towards zero as the disorder potential is increased.  The anticorrelation is clear from the single-impurity results at short distances shown in Fig. \ref{fig:singleimpuritycombinedplot}, but is evidently destroyed by interference as impurity wavefunctions begin to overlap.  

In summary, we find that even for a highly inhomogeneous superconductor, with impurities spread densely throughout the sample, a strong correlation can be seen between the  order parameter (or, equivalently, $I_c$) and the coherence-peak height, but not the spectral gap. This is in fact what is reported in the iron-based superconductor FeTeSe\cite{cho2019strongly}: a large correlation ($r \approx 0.6$) was measured between the coherence-peak height and the quantity $I^2_c R^2_N$,  which in our analysis is essentially equivalent to a strong correlation between the CPH and OP. On the other hand,  there was no observed correlation between the spectral gap and $I^2_c R^2_N$ in experiment.
These results find a natural explanation here from the response of the superconducting condensate to disorder.

\section{Smooth Disorder} 
\label{sec:smoothdisorder}

\begin{figure}[t]
	\centering
	\includegraphics[width=0.5\textwidth]{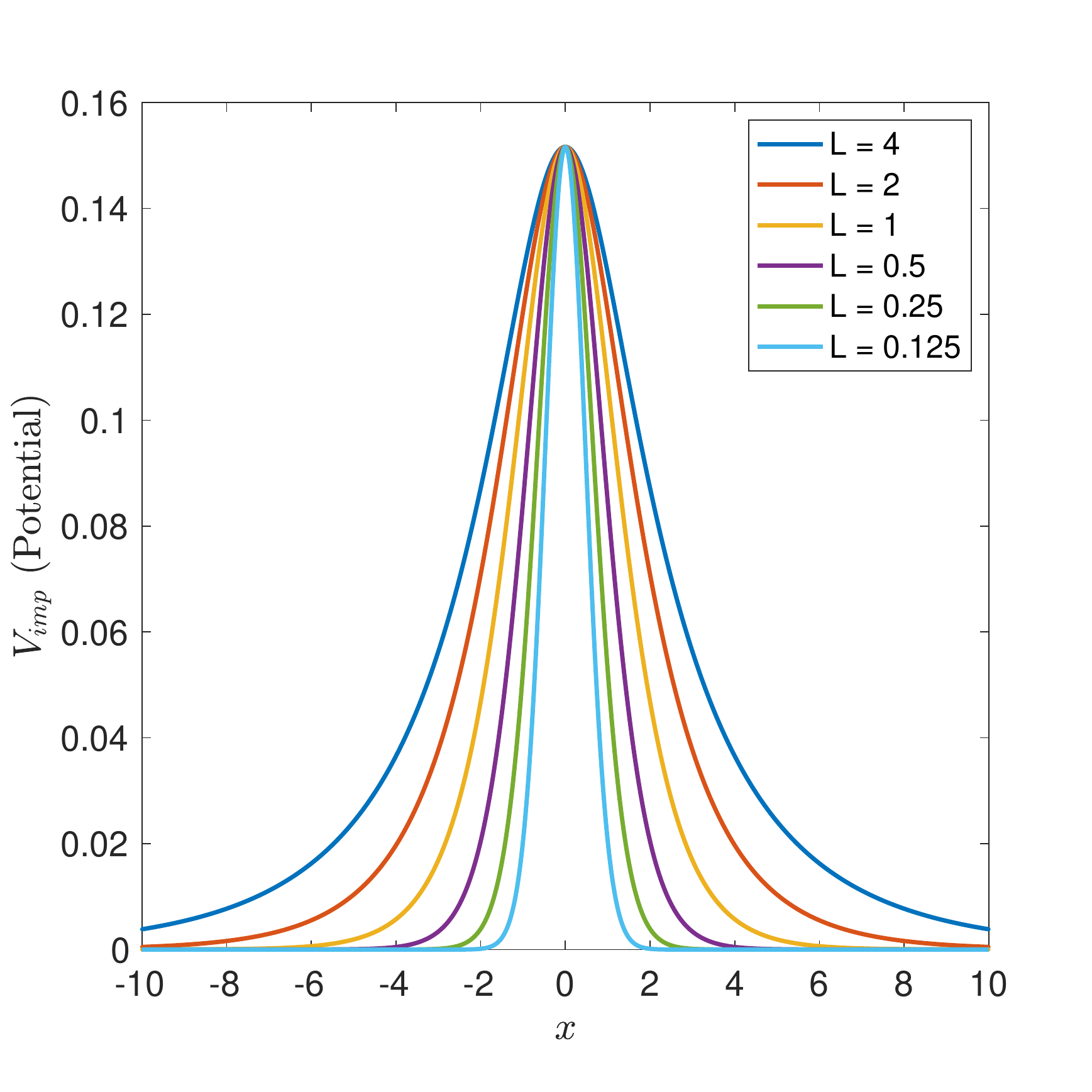}\\
	\caption{Plots of the potential of a single off-plane scatterer vs. the $x$-coordinate under the tuning protocol used in the paper, shown for various values of $L$.  }
	\label{fig:offplanepotentialplot}
\end{figure}

\begin{figure*}[ht]
	\centering
	\includegraphics[width=1.0\textwidth]{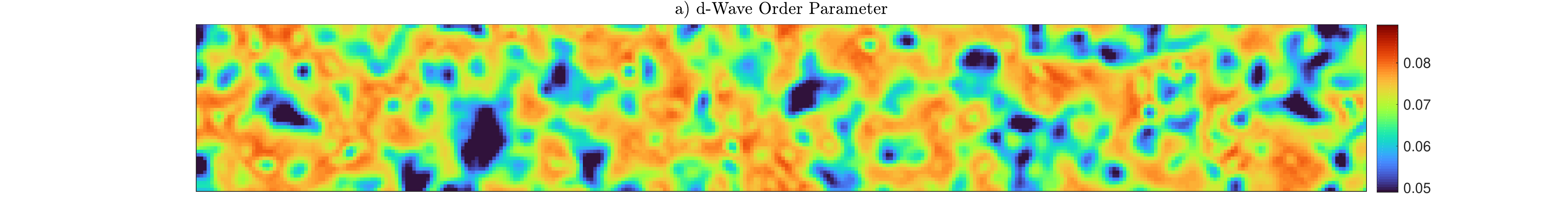} \\
	\includegraphics[width=1.0\textwidth]{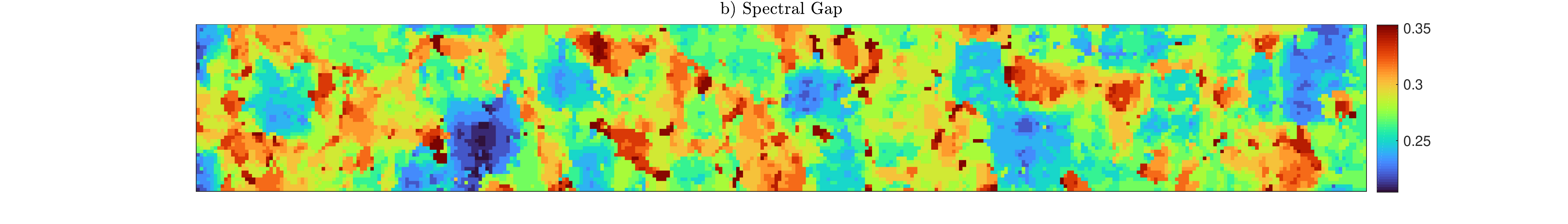}  \\
	\includegraphics[width=1.0\textwidth]{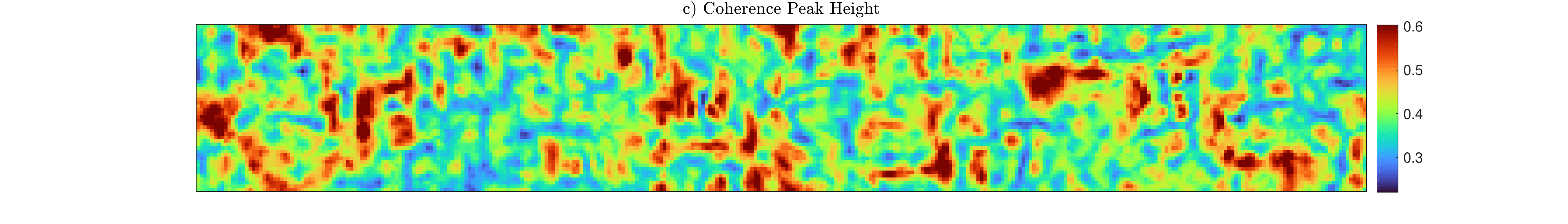}  
	\caption{Plots of (top to bottom) the $d$-wave order parameter, the spectral gap, and the coherence-peak height for a $d$-wave superconductor with smooth disorder with screening length $L = 2$.}
	\label{fig:smoothcombinedplot}
\end{figure*}

\begin{figure*}[t]
	\centering
	\includegraphics[width=0.15\textwidth]{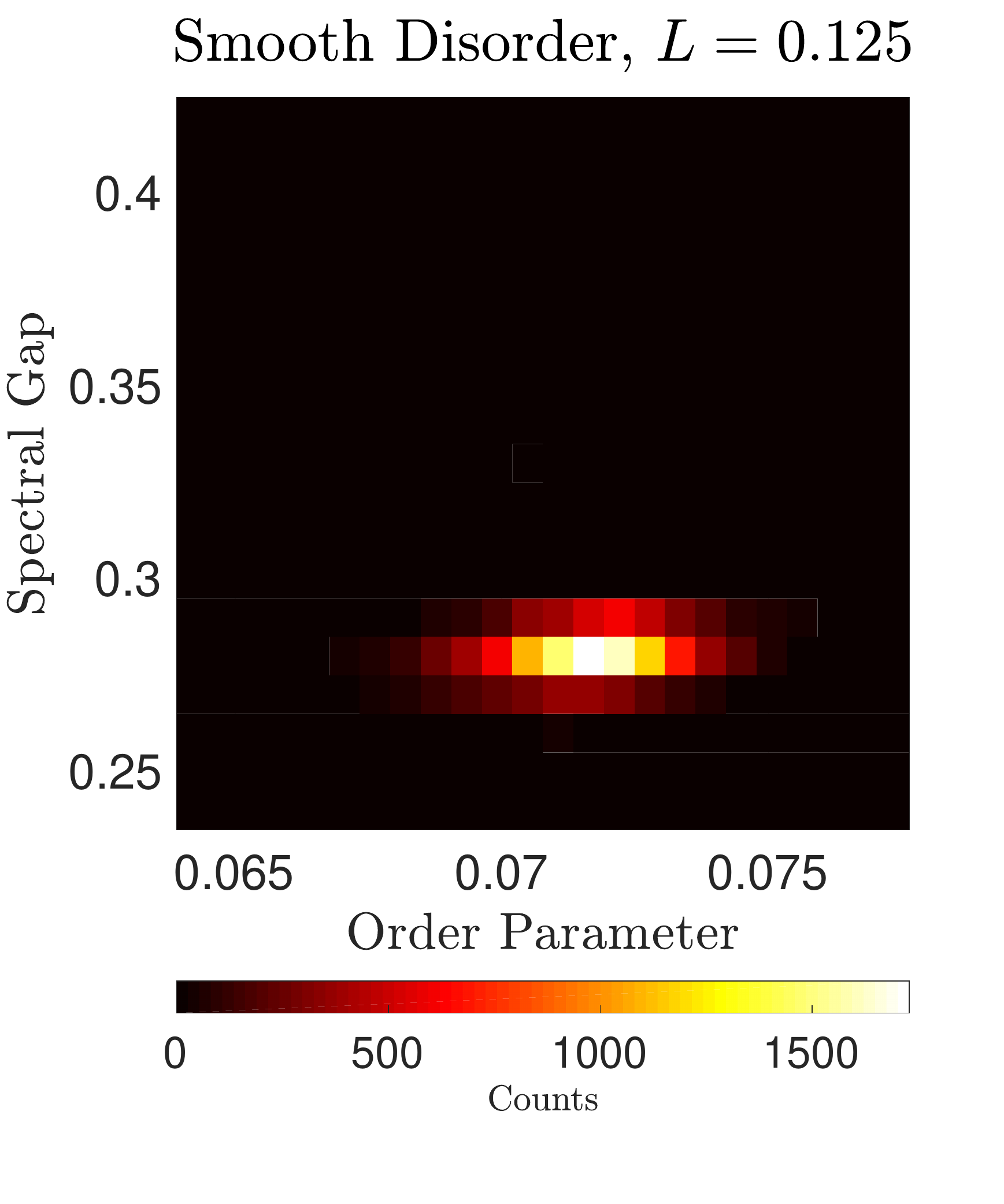}\quad
	\includegraphics[width=0.15\textwidth]{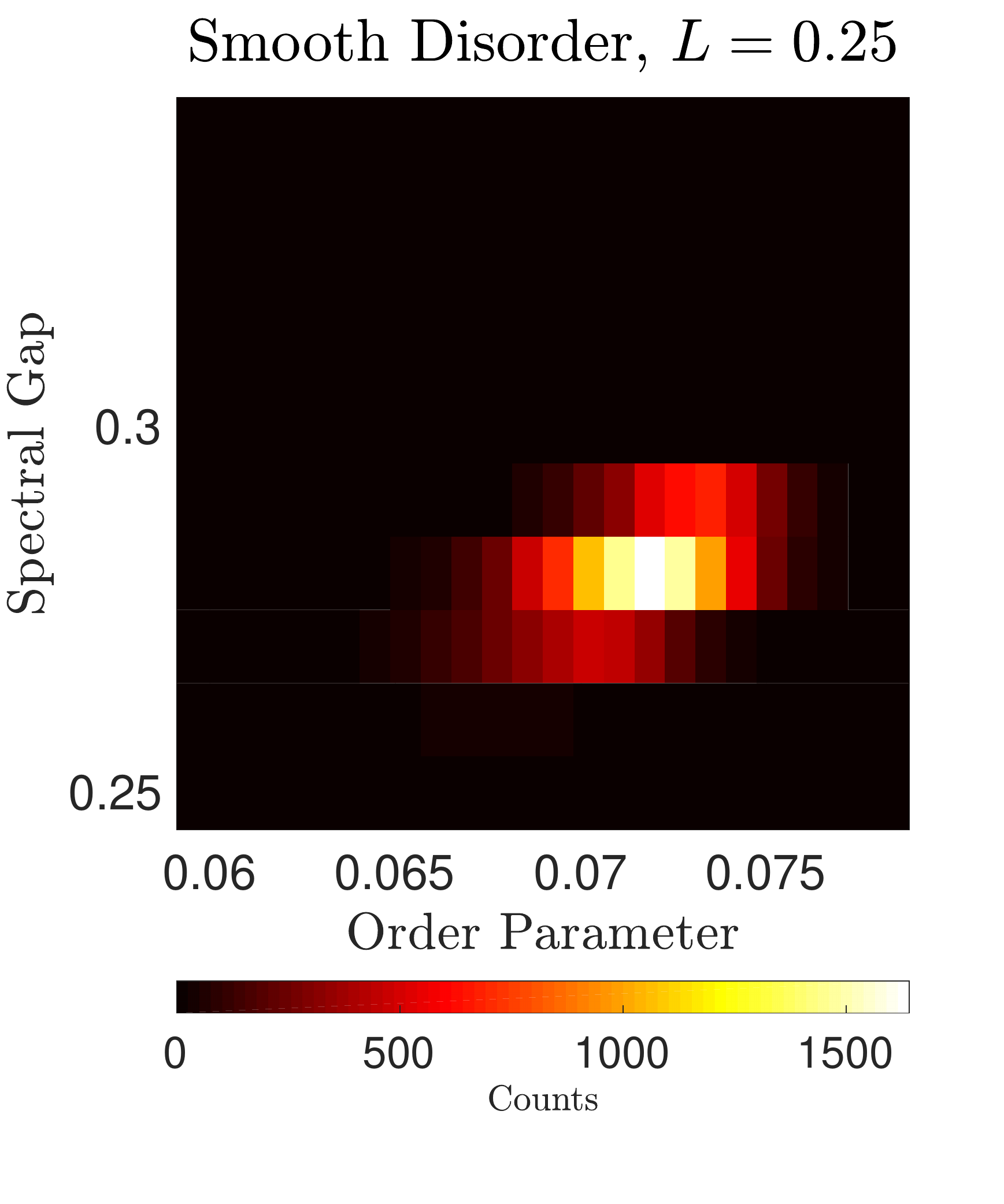}\quad
	\includegraphics[width=0.15\textwidth]{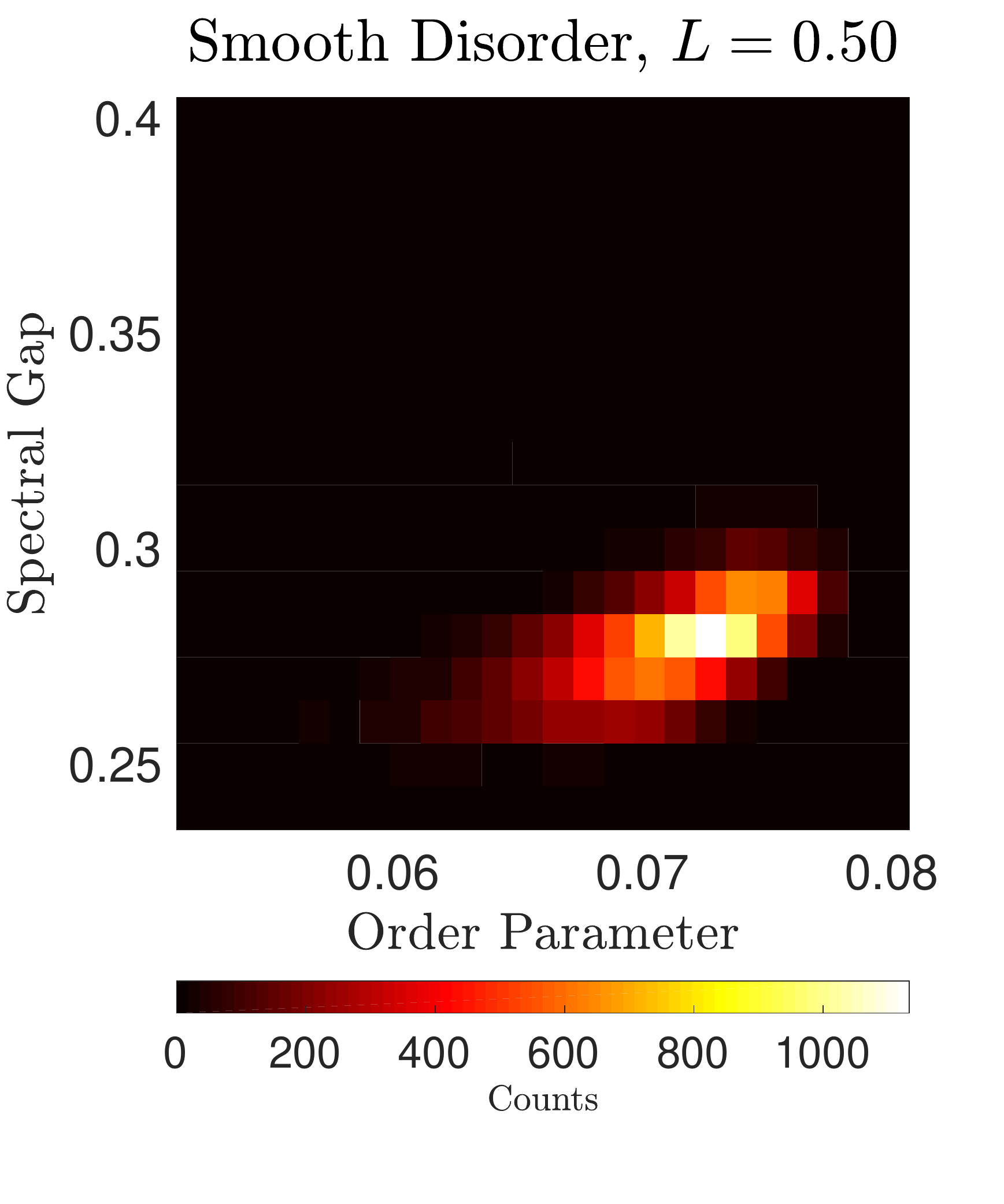}\quad
	\includegraphics[width=0.15\textwidth]{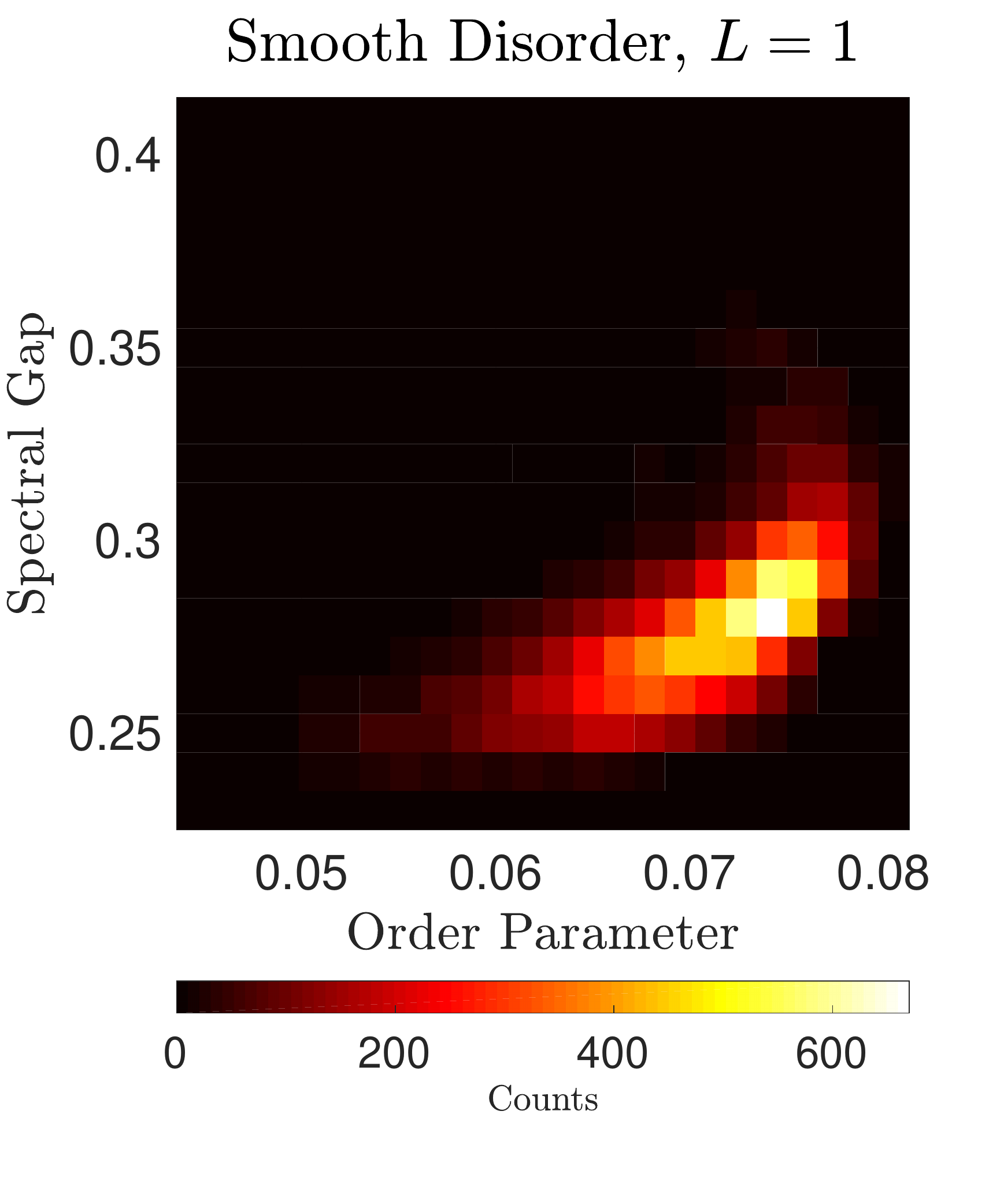}\quad
	\includegraphics[width=0.15\textwidth]{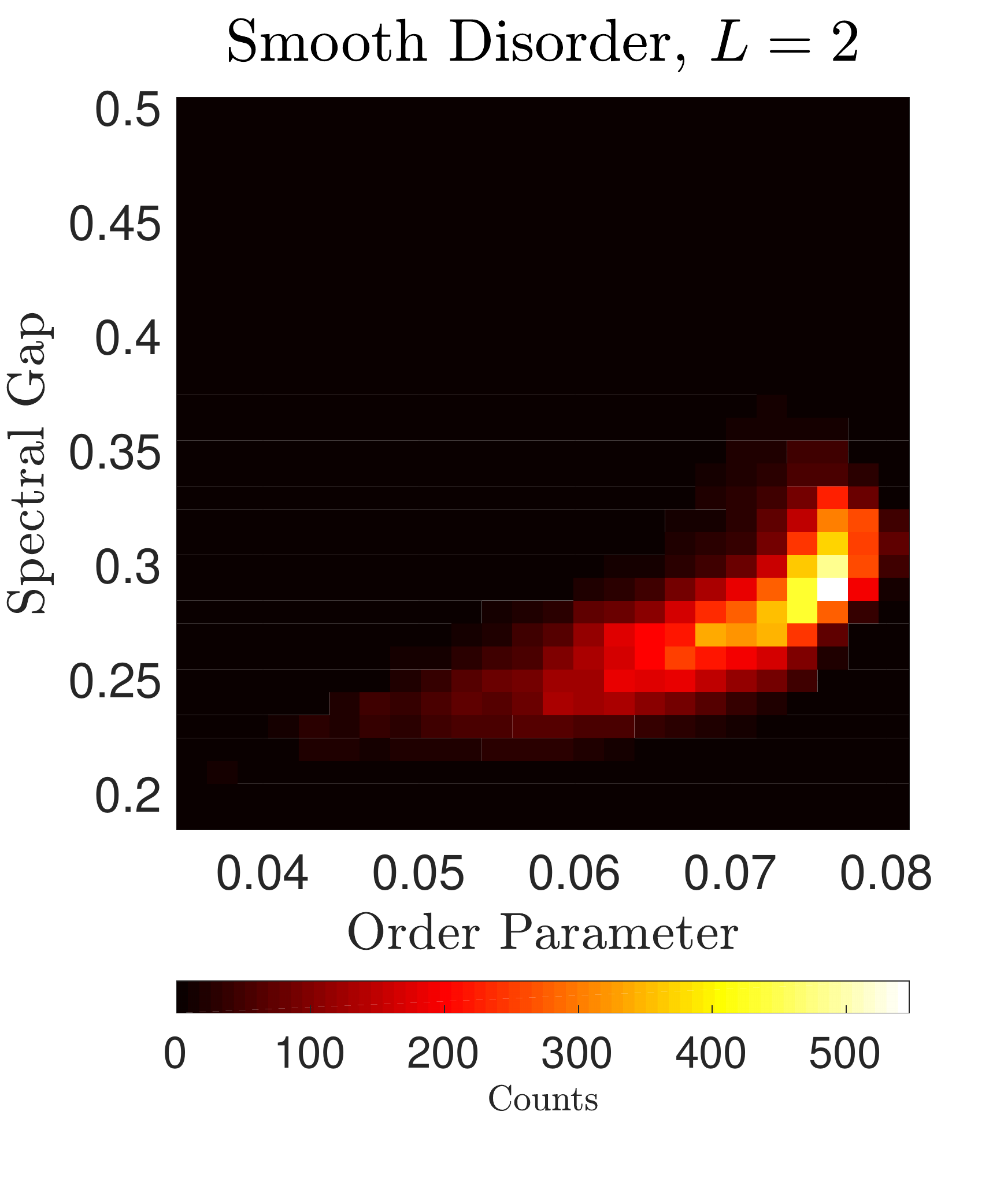}\quad
	\includegraphics[width=0.15\textwidth]{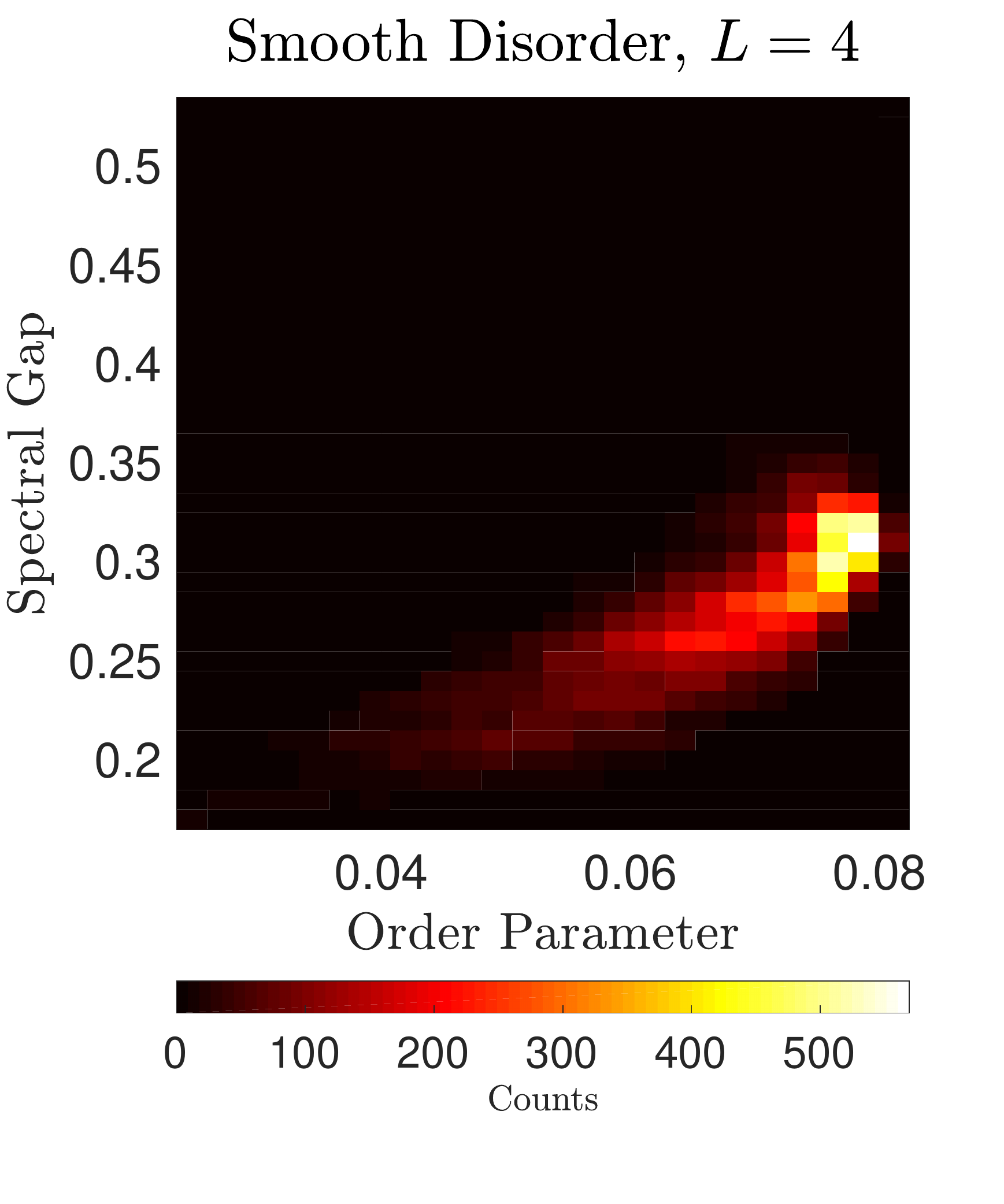}\\
	
	\includegraphics[width=0.15\textwidth]{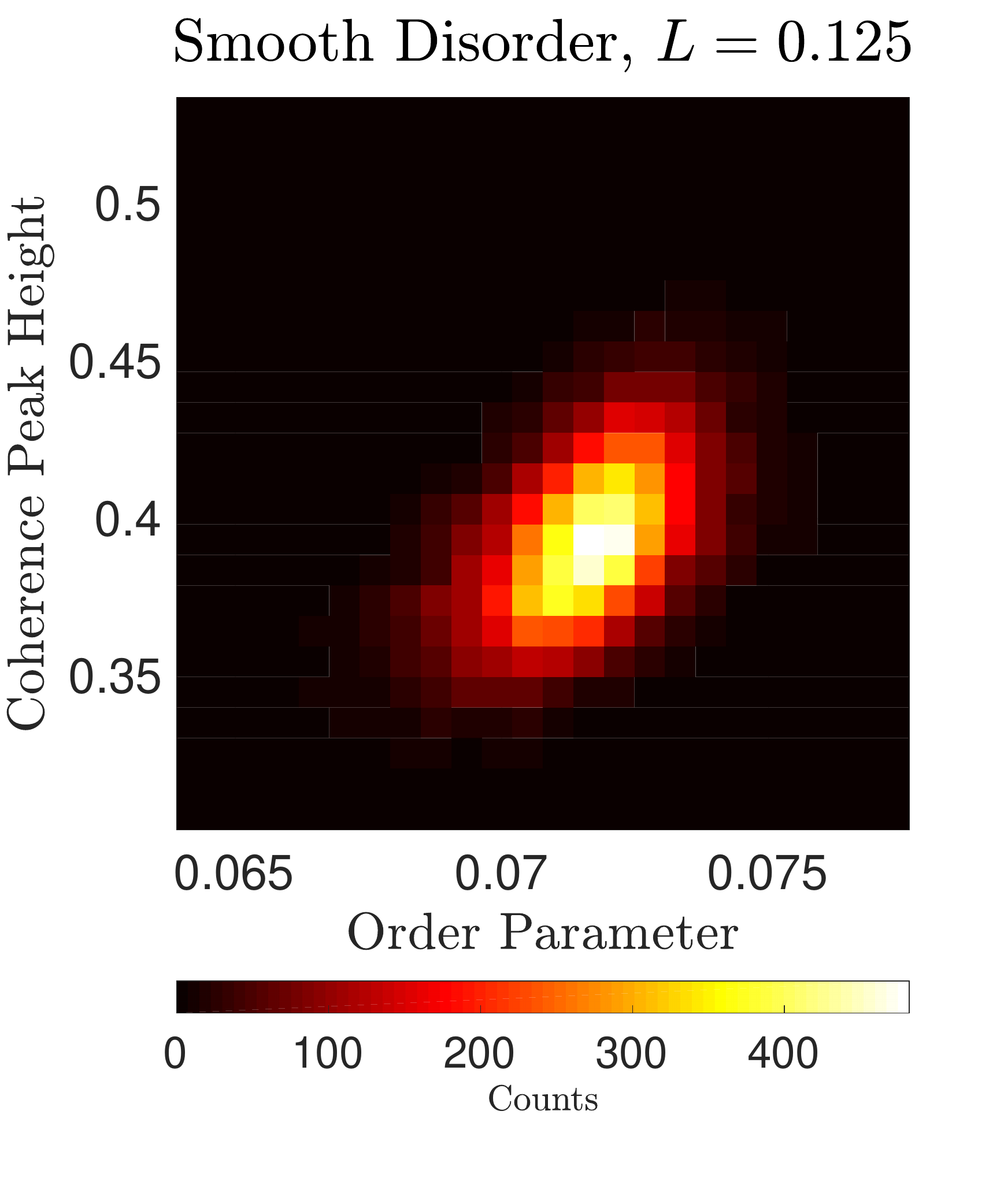}\quad
	\includegraphics[width=0.15\textwidth]{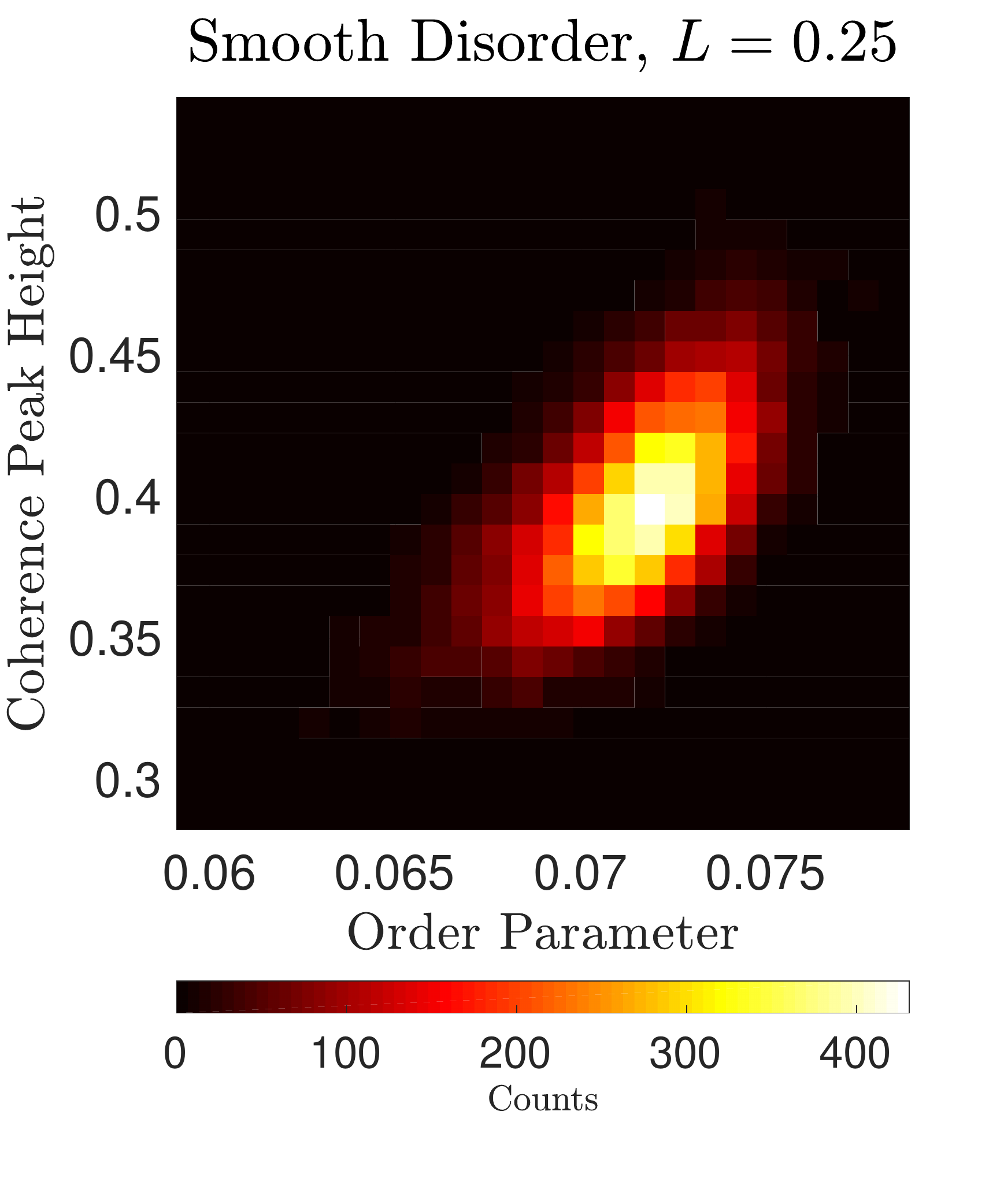}\quad
	\includegraphics[width=0.15\textwidth]{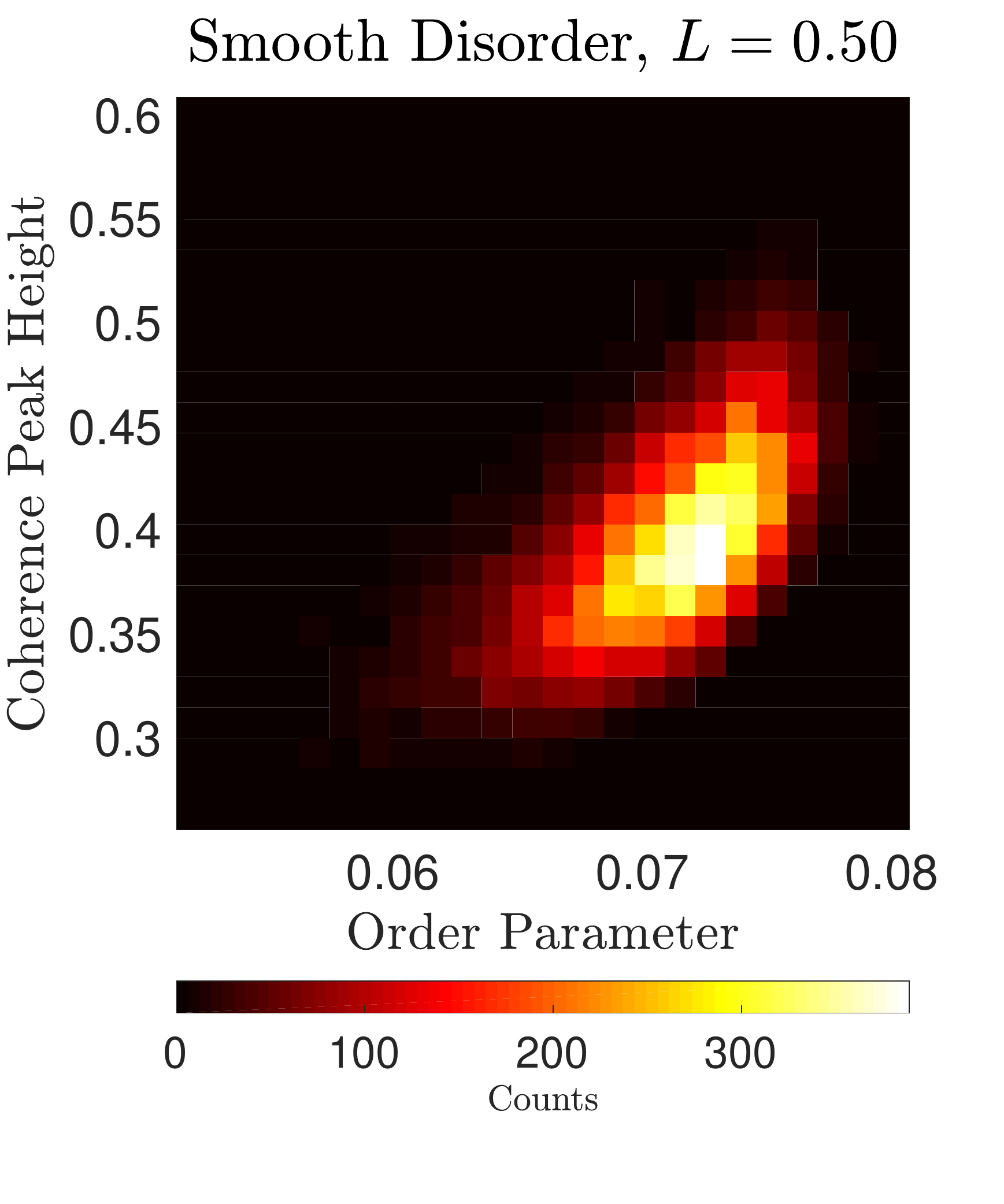}\quad
	\includegraphics[width=0.15\textwidth]{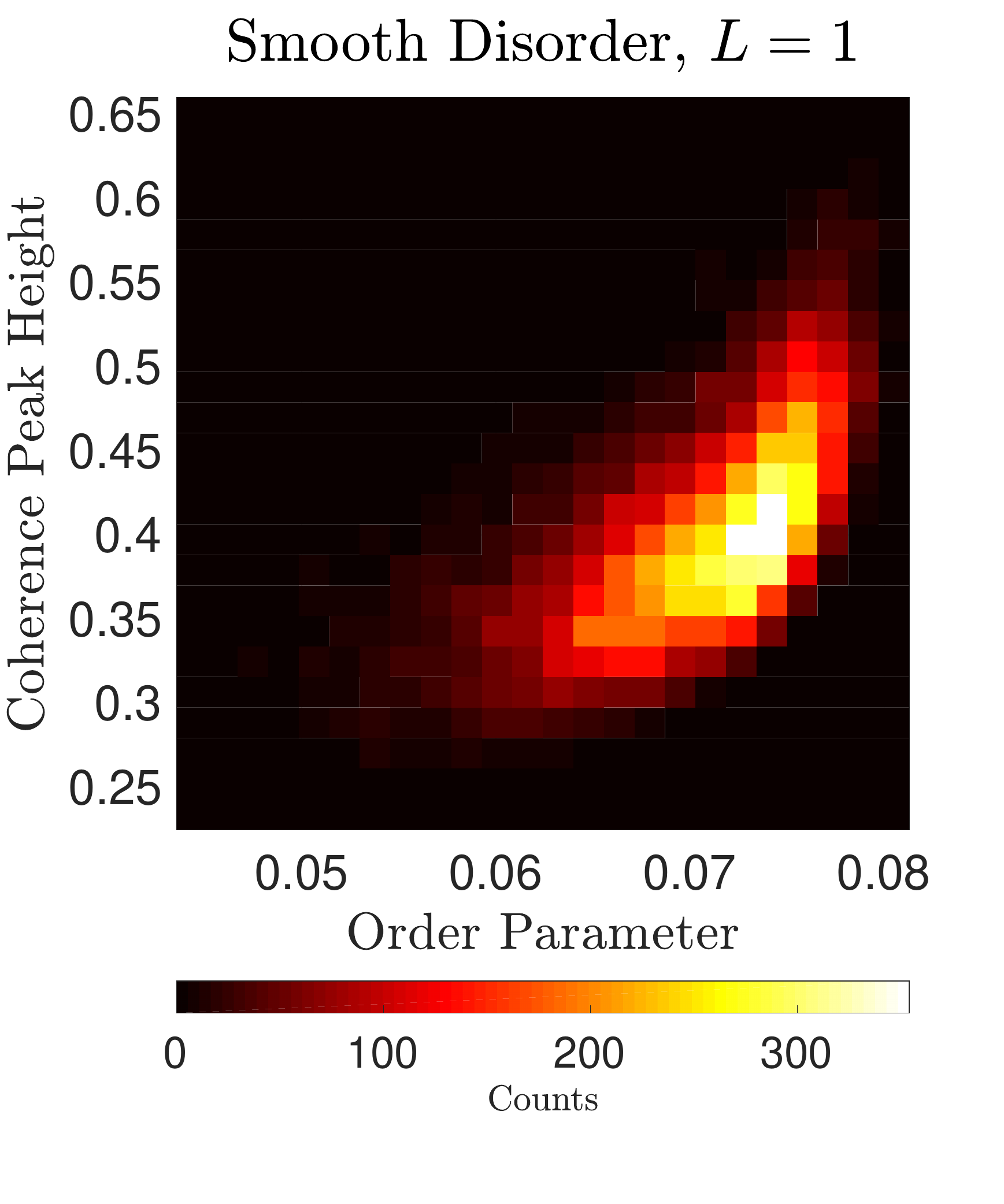}\quad
	\includegraphics[width=0.15\textwidth]{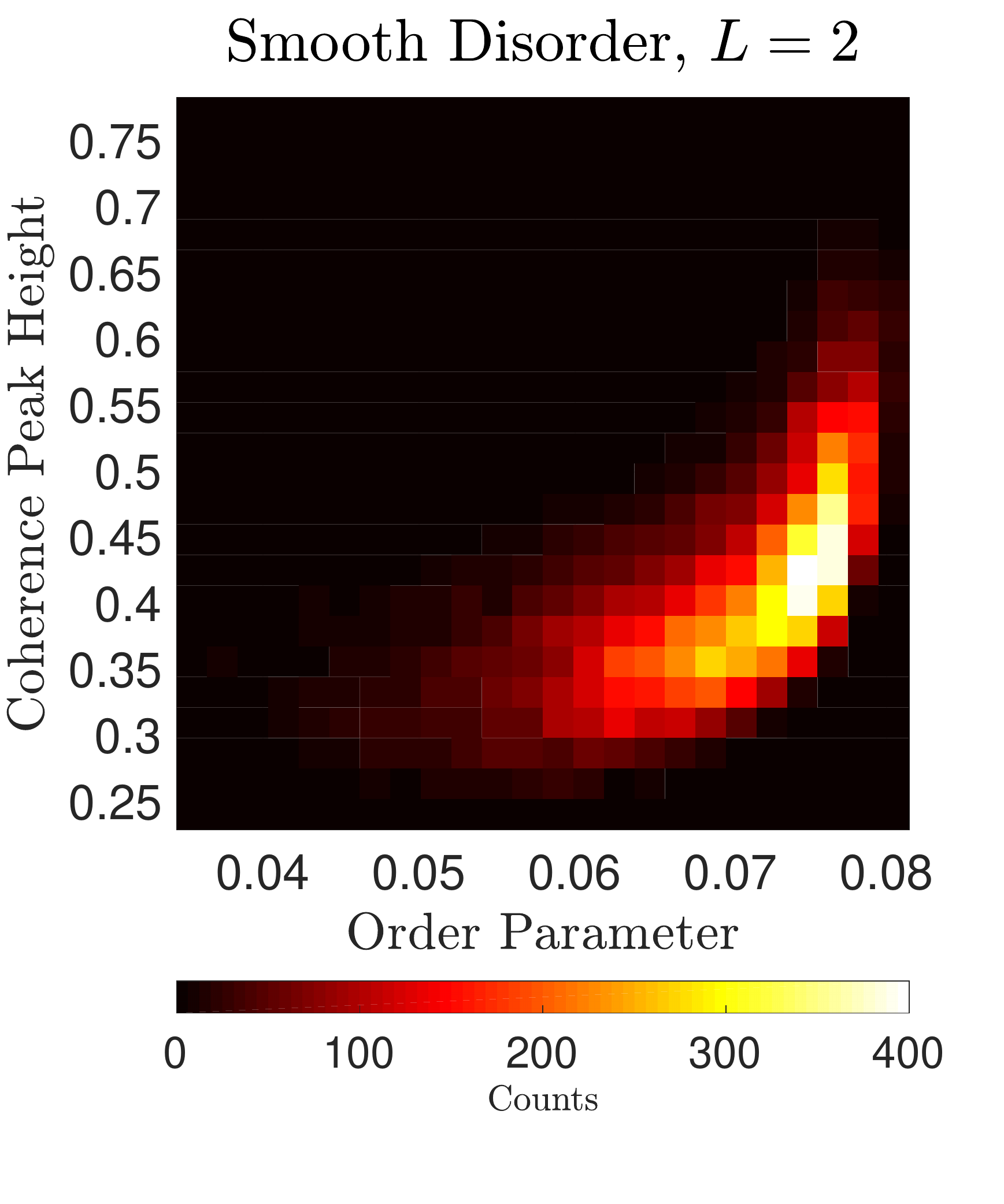}\quad
	\includegraphics[width=0.15\textwidth]{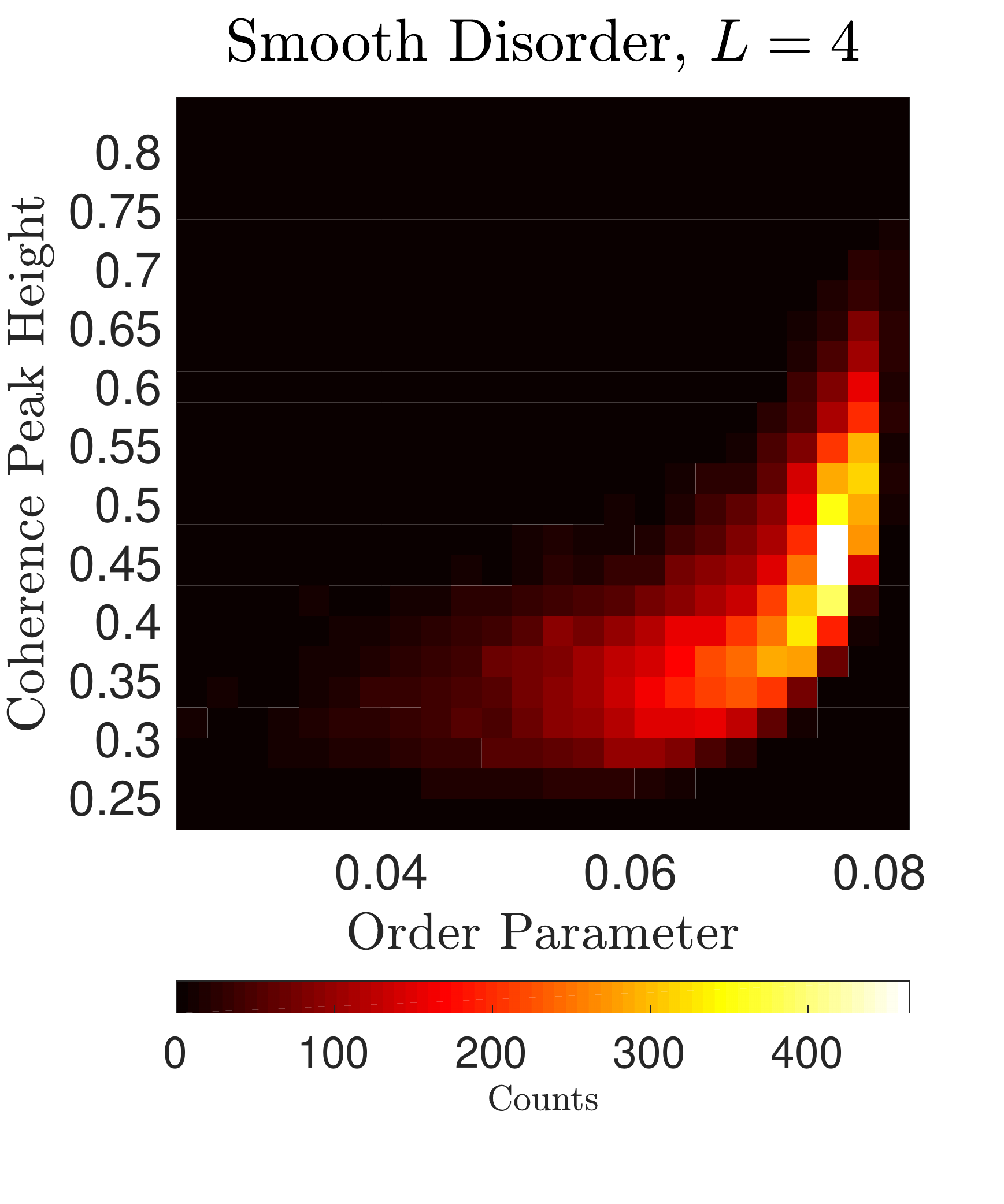}\\
	
	\caption{Two-dimensional histograms between the order parameter and the spectral gap (top row) and the order parameter and the coherence-peak height (bottom row), shown for smooth disorder with varying screening length strength $L$ (left to right). Note that the scales of the $x$- and $y$-axes are not the same as $L$ increases.}
	\label{fig:smoothhistograms}
\end{figure*}

\begin{figure}[t]
	\centering
	\includegraphics[width=0.5\textwidth]{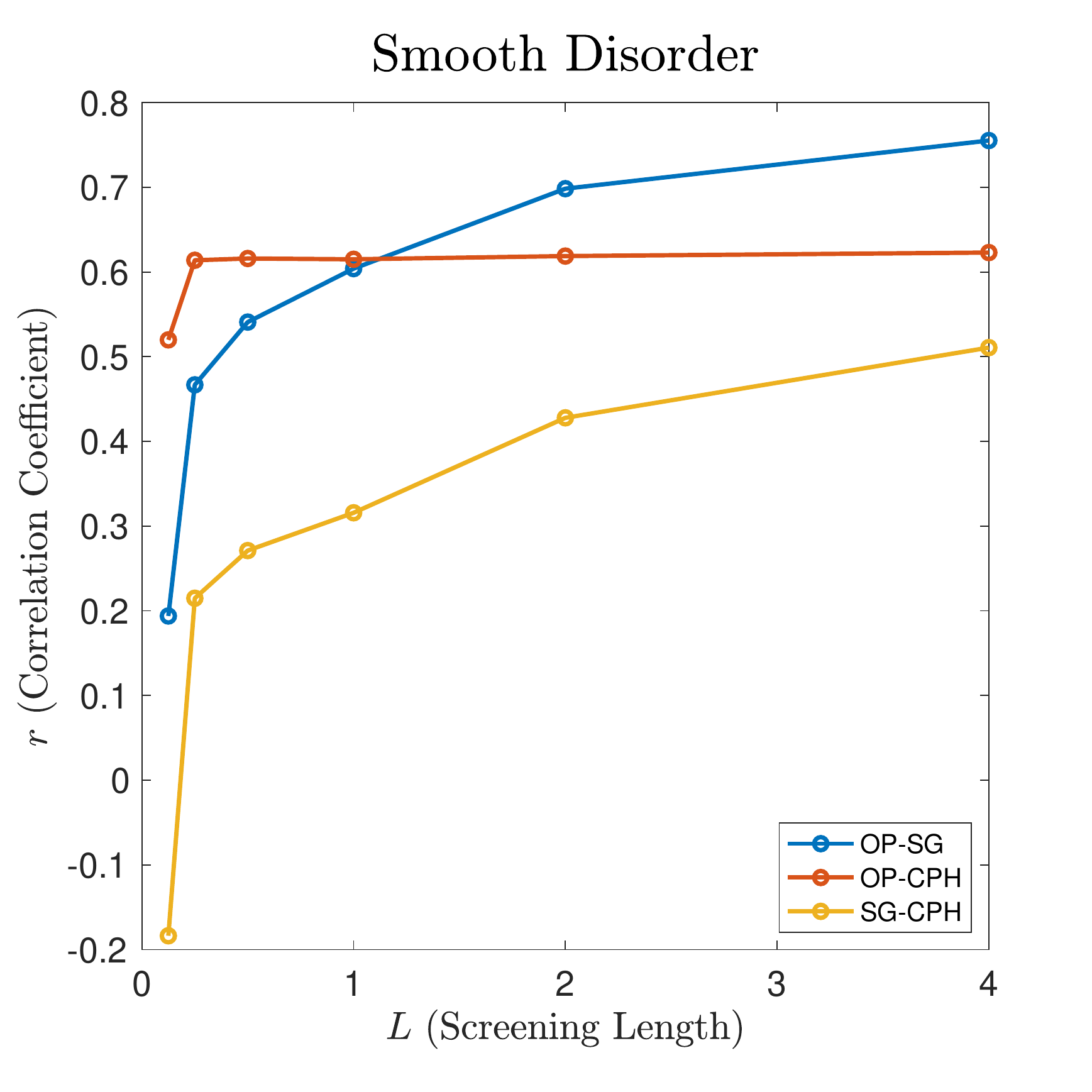}\\
	\caption{Correlation coefficients $r$ for various pairs of quantities as a function of smooth-disorder screening length $L$. }
	\label{fig:smoothcorr}
\end{figure}

Lastly, we consider smooth disorder---\emph{i.e.}, disorder with a length scale larger than the lattice constant. This form of disorder has been hypothesized to be central to cuprates such as BSCCO that host dopants located in the insulating layers away from the copper-oxide planes,\cite{eisaki2004effect,Eisaki2005} although other work suggests that the actual disorder model for BSCCO is more complex.\cite{nunner2005dopant,sulangi2017revisiting} It is, nonetheless, instructive to consider a smooth potential as a point of comparison to the models of Secs.~\ref{sec:pointlike} and \ref{sec:binaryalloy}, which feature atomic-scale variations of the potential. 

Smooth disorder can be understood simply as impurities that generate mostly \emph{forward} scattering. Despite the seeming intractability of this disorder model, disorder-averaged treatments of weak purely forward scatterers in $d$-wave superconductors exist, which allow straightforward conclusions to be drawn as to the strength of pair-breaking and the effects on $T_c$ due to these impurities. It was found that in the weak purely forward-scattering limit, no pair-breaking occurs due to these impurities, similar to Anderson's theorem for $s$-wave superconductors with nonmagnetic scatterers.\cite{zhu2004elastic,graser2007t} It was also found that the suppression of $T_c$ within a purely forward-scattering disorder model is much smaller than that within a pointlike Born scattering model for a given concentration of scatterers.\cite{kee2001effect} Nevertheless, the complicated nature of this form of disorder demands a primarily numerical approach when studying its applications to the cuprates; in particular, the subtle effects of self-consistency in the order parameter are neglected by these analytical approaches. In this paper we study smooth-disorder levels that are beyond the weak-disorder regimes that are accessible by analytical treatments.

We assume that the disorder originates from randomly distributed off-plane dopants that generate screened Coulomb potentials that act as perturbations to the onsite potential.\cite{zhu2004power,nunner2005dopant,nunner2006fourier,sulangi2017revisiting,sulangi2018quasiparticle} We take our model for the smooth disorder potential to be a screened Coulomb potential,
\begin{equation}
V_\mathrm{imp}(j) = \sum_{i\in\{i\}_\mathrm{imp}} \alpha_i V_s   \frac{  e^{-r_{ij}/L}}{r_{ij}},
\label{eq:V}
\end{equation}
where $r_{ij} = \sqrt{|i-j|^2+z_0^2}$ is the distance between  lattice site $j$ and an impurity situated a distance $z_0$ above site $i$. Here, $L$ is the screening length, and $V_s$ governs the strength of the single-impurity potential.  The factor $\alpha_i$ takes the values $\pm 1$ with equal probability, and is introduced to reduce the amount of electron-doping induced by disorder.  We present simulations assuming that the dopants are located a distance $z_0 = 2$ lattice constants away from the CuO$_2$ plane.

The parameter $L$ governs the range of the potential,  with small $L$ corresponding to pointlike impurities. We wish to highlight the influence  of the finite range of impurities to compare to the other cases considered here, but a direct comparison allowing isolation of this effect is not straightforward.  We illustrate one possible comparison in Fig.~\ref{fig:offplanepotentialplot}, which shows the potential plotted within the superconducting plane created by a single impurity. In each case, $V_s$ is chosen to give the same value at $x=0$.  Note that even when $L = 0.25$, the potential is still spread out such that its value on the nearest-neighbor site ($x = \pm1$) is around a third of its value at $x = 0$. It is only at $L = 0.125$ where the potential is close to the pointlike limit.

In the case of sufficiently smooth disorder, $L\gg \xi$, we expect superconductivity to be essentially uniform in the presence of a local chemical potential set by disorder.  In this extreme case, we expect the order parameter and the spectral gap to be well correlated with each other. The approach to this limit can already be seen to some extent in Figure~\ref{fig:smoothcombinedplot}, which shows plots of the $d$-wave component of the order parameter, the spectral gap, and the coherence-peak height for the smooth-disorder case with $L = 2$ and $p = 20\%$. Unlike in the pointlike-disorder cases shown earlier, here there is a visible correlation between the order parameter and the spectral gap. One can make a fairly straightforward match between features belonging to one map and those belonging to another. There is also a visible similarity between the order parameter and the coherence-peak height. However, the features seen in  both the spectral gap and the coherence-peak height maps have more structure than those in the order-parameter map, and one can see significant fluctuations of the peak height in regions where the order parameter is large and uniform.   We believe this represents an impurity interference effect in this limit: while the order parameter averages over a region of order $\xi$, the spectral gap and coherence peak height are determined by the interference of the nearby impurity wavefunctions. When smooth disorder is present, quasiparticle interference at energies near the gap edge is dominated by scattering wavevectors $\mathbf{q}$ whose magnitudes are parametrically smaller than $2k_F$.\cite{nunner2006fourier,sulangi2017revisiting,sulangi2018quasiparticle} This nevertheless gives rise to modulations in the LDOS whose length scale is set by $q^{-1}$, and consequently to the real-space variations clearly visible in the coherence-peak height plots.  

We show the two-dimensional histograms between the order parameter and both the spectral gap and the coherence-peak height for increasing screening lengths $L$ in Fig.~\ref{fig:smoothhistograms}. The plots for $L=0.125$ are similar to the pointlike case discussed in Sec.~\ref{sec:pointlike}:  the spectral gap is weakly correlated with the order parameter, but there is a strong positive correlation between the order parameter and coherence peak height.   As in Fig. \ref{fig:smoothcombinedplot}, the significant difference from the pointlike case is that, as $L$ increases, a positive correlation develops between the order parameter and the spectral gap. The relationship between the two quantities is approximately linear, but it shows a slight upwards curvature at large $L$.  This curvature is much more pronounced in the relationship between the order parameter and the coherence peak height.  This upturn is a reflection of what we have observed in Fig.~\ref{fig:smoothcombinedplot}, namely that there are large variations of the coherence peak height in areas where the order parameter is uniform and large.  Such regions are approximately perfectly clean $d-$wave superconductors, so that the coherence peak height is in principle logarithmically infinite.

The correlation coefficients $r$, shown in Fig.~\ref{fig:smoothcorr}, confirm the observations we have made from Figs.~\ref{fig:smoothcombinedplot} and~\ref{fig:smoothhistograms}. $r$ between the order parameter and the spectral gap is as high as 0.76 when $L = 4$, decreasing monotonically as $L$ is lowered, and even when $L = 0.25$, $r \approx 0.47$, much larger than what can be seen in the dilute-impurity and binary-alloy models we have encountered in the earlier sections. However, the OP-SG correlation goes down sharply for $L = 0.125$, rapidly approaching the pointlike limit, with a small $r \approx 0.2$.  These results suggest that the crossover between the smooth limit and pointlike limit is complicated and can depend on the quantity in question, again because the atomic scale impurity wavefunctions influence the LDOS-derived quantities more than the order parameter.

Interestingly, $r$ between the order parameter and the coherence-peak height is almost a constant function of $L$ for $0.25 \leq L \leq 4$, with its value around $r \approx 0.6$. This $r$ is also markedly higher than the corresponding correlation coefficients we have found for weak binary-alloy disorder. In contrast to pointlike disorder, we find an overall positive correlation between the spectral gap and the coherence-peak height for most of the range of $L$, which varies from around 0.2 when $L = 0.25$ to 0.5 when $L = 4$. When $L = 0.125$, however, the SG-CPH correlation becomes negative, similar to the pointlike disorder cases studied earlier. 

\section{Discussion and Conclusion}

In this paper, we have examined the correlations between various experimentally measurable quantities, and we find that the observed lack of correlation between the true order parameter (as measured from $I_c$ maps) and the spectral gap (as obtained from $dI/dV$ measurements) can be explained simply as the effect of disorder. When one has signficant levels of  pointlike disorder, consistent, e.g., with disorder in overdoped cuprates, the correlation is weak for a wide range of disorder strengths. In contrast, the correlation is found to be strong when disorder is smooth in the sense that the impurity range or disorder correlation length is comparable to or larger than the coherence length $\xi_0$. We also find a fairly prominent correlation between the order parameter and the coherence-peak height, which can be attributed to spectral-weight transfer due to the presence of weak impurities.

We do not intend to claim that disorder alone is responsible for these effects---clearly the phenomenology of BSCCO demands that interaction effects that act inhomogeneously throughout the system (\emph{e.g.}, a spatially varying scattering rate) be present; these are not taken into consideration in our models. Our main point is that disorder could account for a good part of the mystery of why the order parameter and the spectral gap are not necessarily correlated with each other. This explanation is founded on the observation that $I_c$ maps are a near-perfect proxy observable for the superconducting order parameter---not the superfluid density---and once this is taken into consideration, the discrepancy between the two sets of quantities arises as a simple consequence of the reorganization of spectral weight in the presence of disorder. In fact, for FeTeSe, where interaction effects may not be as  important as in the cuprates, the phenomenology contained in  binary alloy  models provides a surprisingly comprehensive explanation for all of the correlations (or the lack thereof) seen between various pairs of experimental measurables.

For the cuprates, on the other hand, the situation is murkier. There is as yet no published systematic analysis establishing definitively the sort of correlation that exists between $I_c$ and the spectral gap in the cuprates. However, our results can shed light on possible explanations should a strong correlation (or the absence thereof) be found between these two quantities in experiment. If the $I_c$ maps and the $dI/dV$ spectral-gap maps are highly correlated (\emph{i.e.}, $r > 0.5$) with each other, then the smooth-disorder model discussed previously provides a minimal explanation that accounts for this agreement.  The observed distribution of spectral gap values in STM studies on BSCCO  precludes an explanation in terms of smooth disorder in the extreme limit $L\gg \xi$, however\cite{nunner2005dopant,nunner2006fourier,sulangi2017revisiting}, so answers will unfortunately depend on details.   

If on the other hand there is a lack of correlation of  the order parameter and the spectral gap, any one of the pointlike models considered in this paper is a likely candidate to explain this effect. The absence of a strong correlation would suggest too that if the disorder in the cuprates were due to off-plane dopants, then these are in the well-screened limit such that they  may be treated as  pointlike scatterers. It is known however that STS experiments on BSCCO find a strong \emph{anticorrelation} between the spectral gap and the coherence-peak height.\cite{lang2002imaging,fang2006gap,alldredge2008evolution} From what we have seen in the models we have considered, this can be partially explained by pointlike disorder (weak dilute impurities, binary-alloy disorder, and off-plane disorder with very small potential range), but not by smooth disorder, which gives rise to a positive correlation instead. It has previously been argued that it is possible to account for this anticorrelation using a phenomenological model of small patches where pairing is enhanced or suppressed embedded within a region with a spatially uniform $d$-wave gap\cite{fang2006gap}; however, this treatment leaves unanswered the question of why these spectral gap ``swimming pools'' or ``plateaus" form at all. It has been suggested\cite{fang2006gap,nunner2005dopant} that these pools may arise naturally as a consequence of disorder, perhaps in the pointlike limit.  It is intriguing that the negative SG-CPH correlation seems to indeed emerge from finite disorder models, consistent with the ``plateau/pool" picture.

However,  the suppression of the coherence-peaks within regions with large spectral gaps is an effect that appears to be beyond the minimal disorder-based models we have considered, since what is seen in experiment is not merely the suppression of the coherence peak within large-spectral gap regions, but a concurrent \emph{broadening} of the spectra. It is likely that this broadening is due to inelastic scattering, which ensures that large-spectral gap regions are broadened much more than small-spectral gap regions, driven by ``local doping'' wherein large-spectral gap and small-spectral gap regions behave similarly to underdoped and overdoped cuprates on average, respectively. This strong-coupling explanation is supported by STS studies which find that a large scattering rate is necessary to account for the suppression of the coherence-peak height in these large-spectral gap regions.\cite{alldredge2008evolution} One can in fact model this anticorrelation phenomenologically with a spatially dependent pairing interaction $V(\mathbf{r}, \mathbf{r'})$ and scattering rate $\eta(\mathbf{r})$, as Graham and Morr had already previously considered within a gap-disorder-only model.\cite{graham2019josephson} However, explaining why both the pairing interaction and the scattering rate are necessarily spatially correlated with each other requires a microscopic treatment that goes beyond our simple, mean-field-based disorder-only model. We hope in any case that the results shown in this paper prove useful to the interpretation of the latest STM experiments on unconventional superconductors.

\begin{acknowledgments}
	
M.A.S received partial support from the IARPA SuperTools program. M.A.S. and P.J.H. received partial support from NSF-DMR-1849751.  W.A.A. acknowledges support by the Natural Sciences and Engineering Research Council (NSERC) of Canada.
 
\end{acknowledgments}

\appendix

\section{Correlations Between the Order Parameter and the Critical Current}

\begin{figure*}[ht]
	\centering
	\includegraphics[width=1.0\textwidth]{dwaveorderparam_point_full50x1000_V=025_020pct".pdf} \\
	\includegraphics[width=1.0\textwidth]{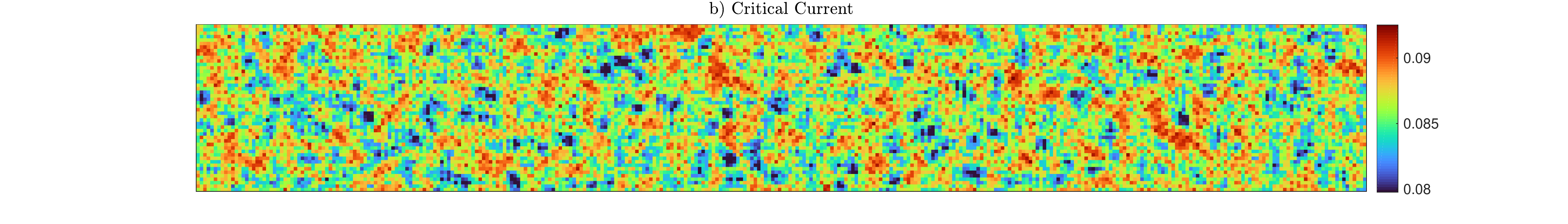}  \\
	\includegraphics[width=1.0\textwidth]{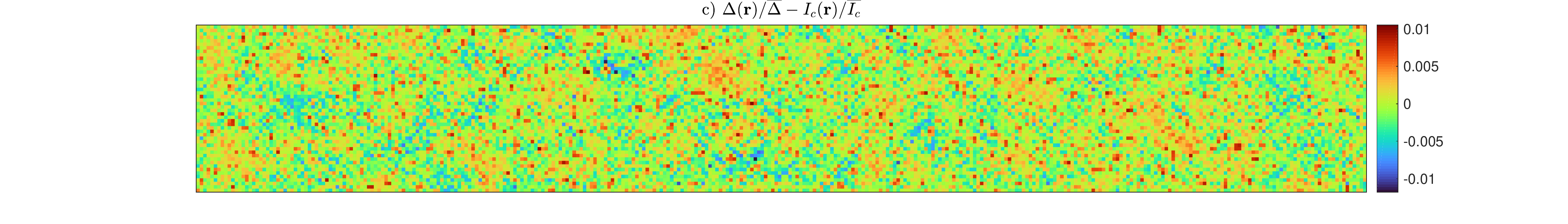}  \\
	\caption{Plots of a) the $d$-wave order parameter, b) the critical current, and c) the normalized difference between a) and b) (\emph{i.e.}, $\Delta(\mathbf{r}))/\overline{\Delta} - I_c(\mathbf{r})/\overline{I_c})$ for a $d$-wave superconductor with weak pointlike impurities with strength $V = 0.25$ and concentration $p = 20\%$.}
	\label{fig:pointlikeopvscurrent}
\end{figure*}

\begin{figure*}[ht]
	\centering
	\includegraphics[width=1.0\textwidth]{dwaveorderparam_binaryalloy_full50x1000_V=0250_050pct".pdf} \\
	\includegraphics[width=1.0\textwidth]{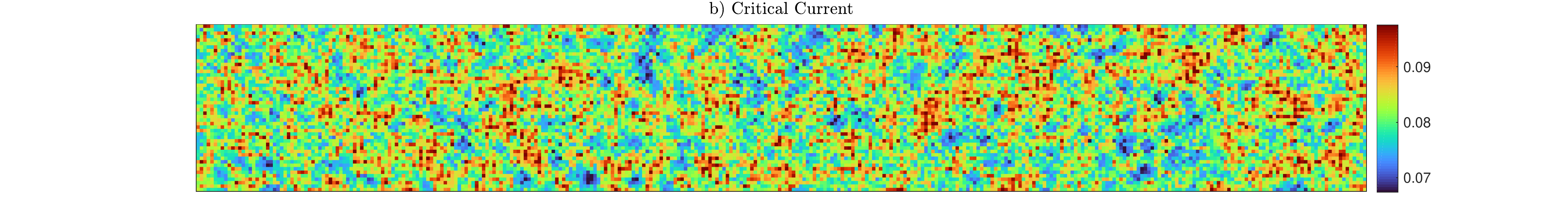}  \\
	\includegraphics[width=1.0\textwidth]{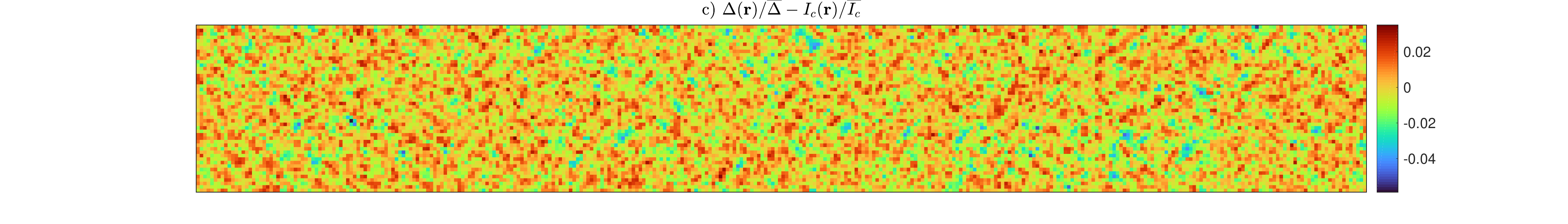}  \\
	\caption{Plots of a) the $d$-wave order parameter, b) the critical current, and c) the normalized difference between a) and b) (\emph{i.e.}, $\Delta(\mathbf{r}))/\overline{\Delta} - I_c(\mathbf{r})/\overline{I_c})$t for a $d$-wave superconductor with binary-alloy disorder with strength $V_b = 0.250$.}
	\label{fig:binaryalloyopvscurrent}
\end{figure*}

\begin{figure*}[ht]
	\centering
	\includegraphics[width=1.0\textwidth]{dwaveorderparam_smooth_full50x1000_V_amp=050_Lscreen=2_z=2_020pct".pdf} \\
	\includegraphics[width=1.0\textwidth]{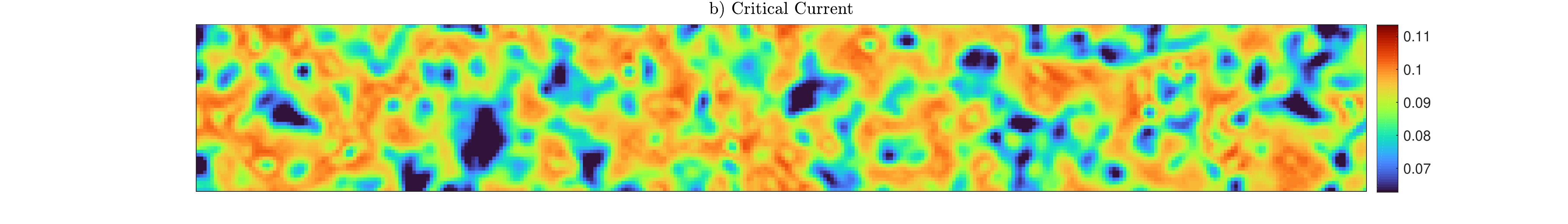} \\
	\includegraphics[width=1.0\textwidth]{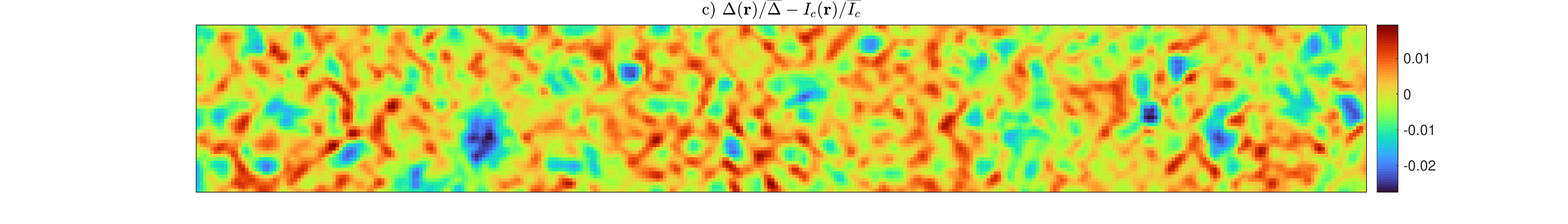} \\
	\caption{Plots of a) the $d$-wave order parameter, b) the critical current, and c) the normalized difference between a) and b) (\emph{i.e.}, $\Delta(\mathbf{r}))/\overline{\Delta} - I_c(\mathbf{r})/\overline{I_c})$ for a $d$-wave superconductor with smooth disorder with screening length $L = 2$.}
	\label{fig:smoothopvscurrent}
\end{figure*}

\begin{figure*}[ht]
	\centering
	\includegraphics[width=.3\textwidth]{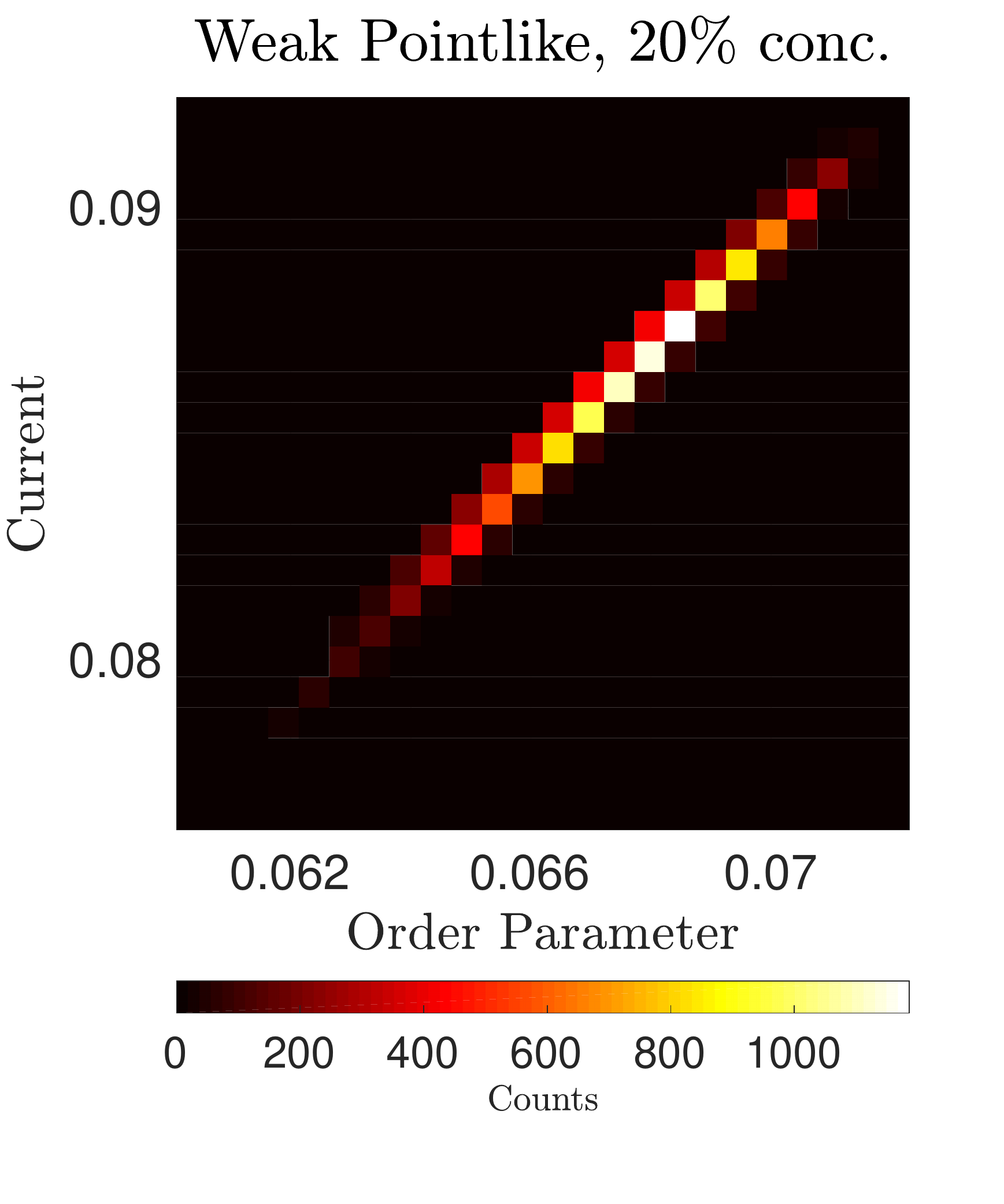}\quad
	\includegraphics[width=.3\textwidth]{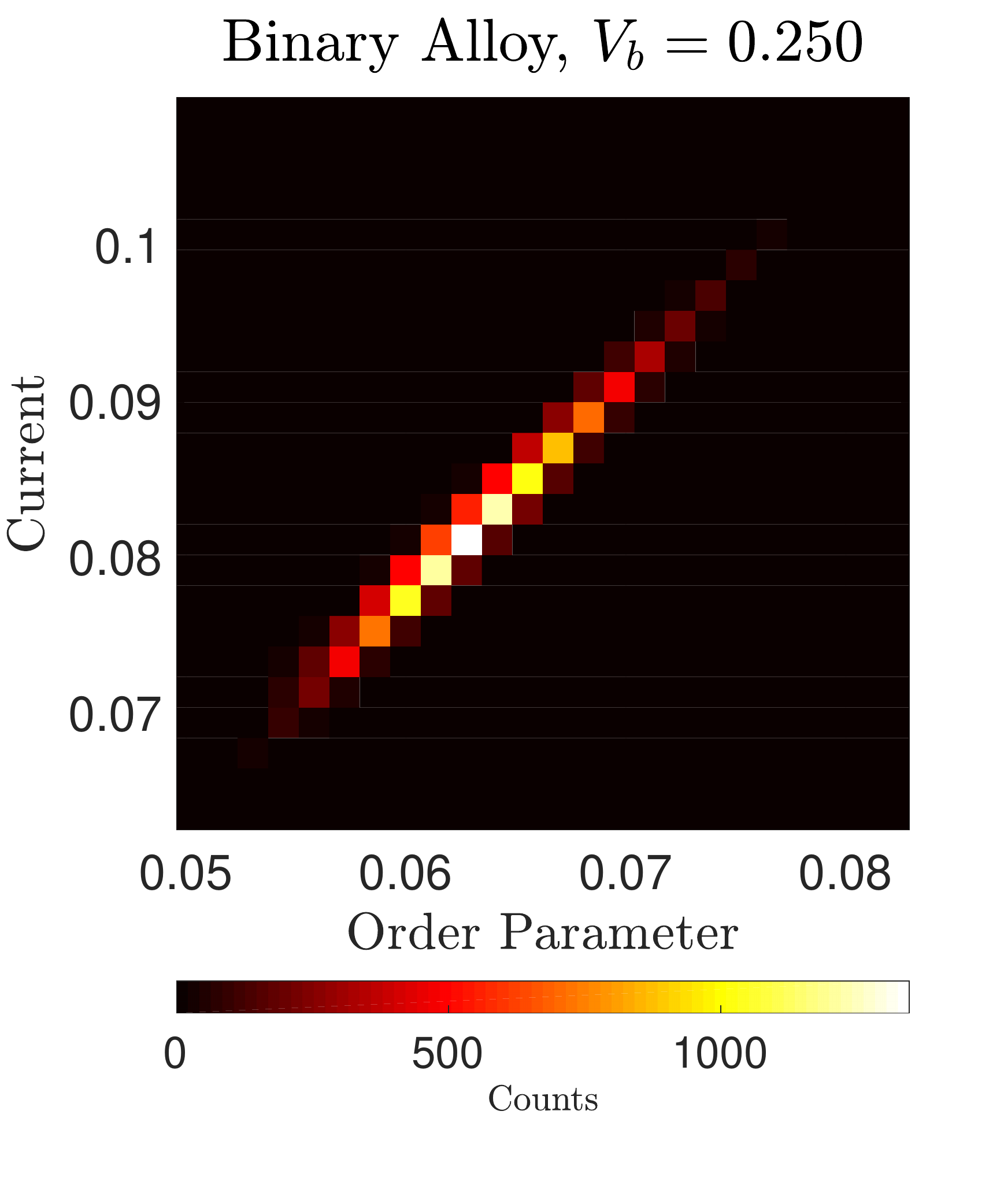}\quad
	\includegraphics[width=.3\textwidth]{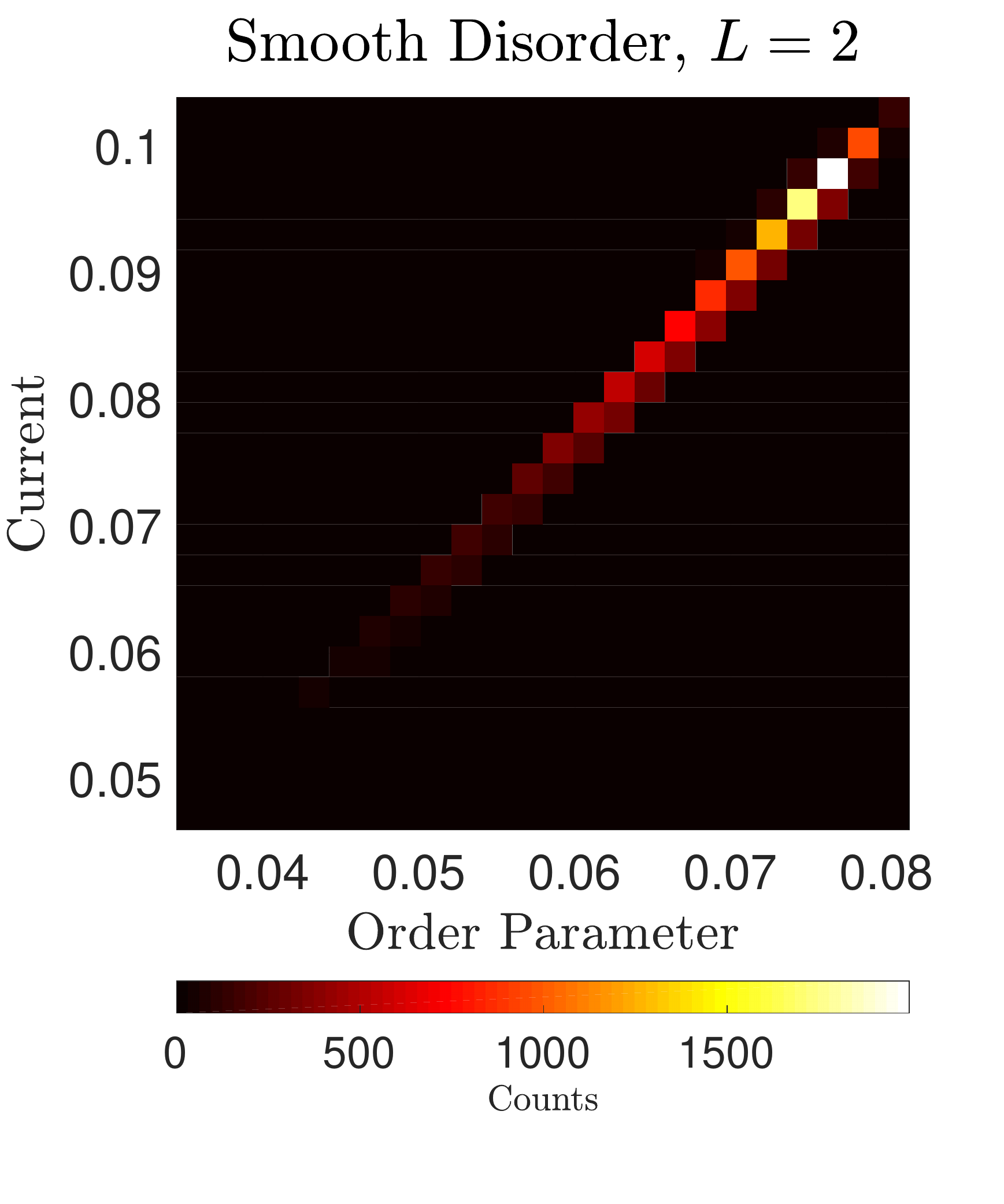}
	\caption{Two-dimensional histograms between the order parameter and $I_c$, shown for three different disorder types corresponding to those plotted in Figs.~\ref{fig:pointlikecombinedplot},~\ref{fig:binaryalloycombinedplot}, and~\ref{fig:smoothcombinedplot}. The values of $r$ for these plots are $0.9958$, $0.9893$, and $0.9984$, respectively.}
	\label{fig:gapcurrenthistograms}
\end{figure*}

\begin{figure*}[ht]
	\centering
	\includegraphics[width=.3\textwidth]{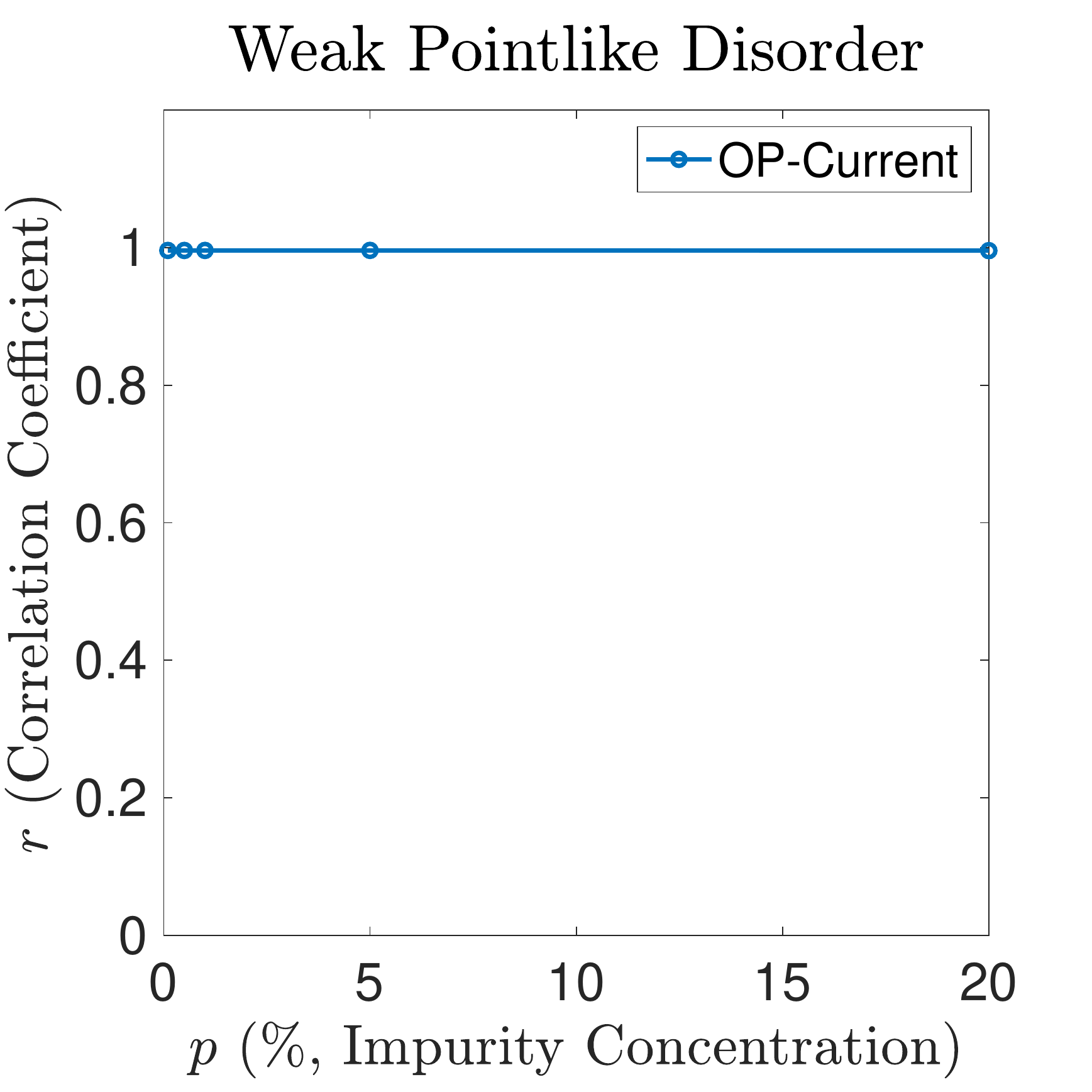}\quad
	\includegraphics[width=.3\textwidth]{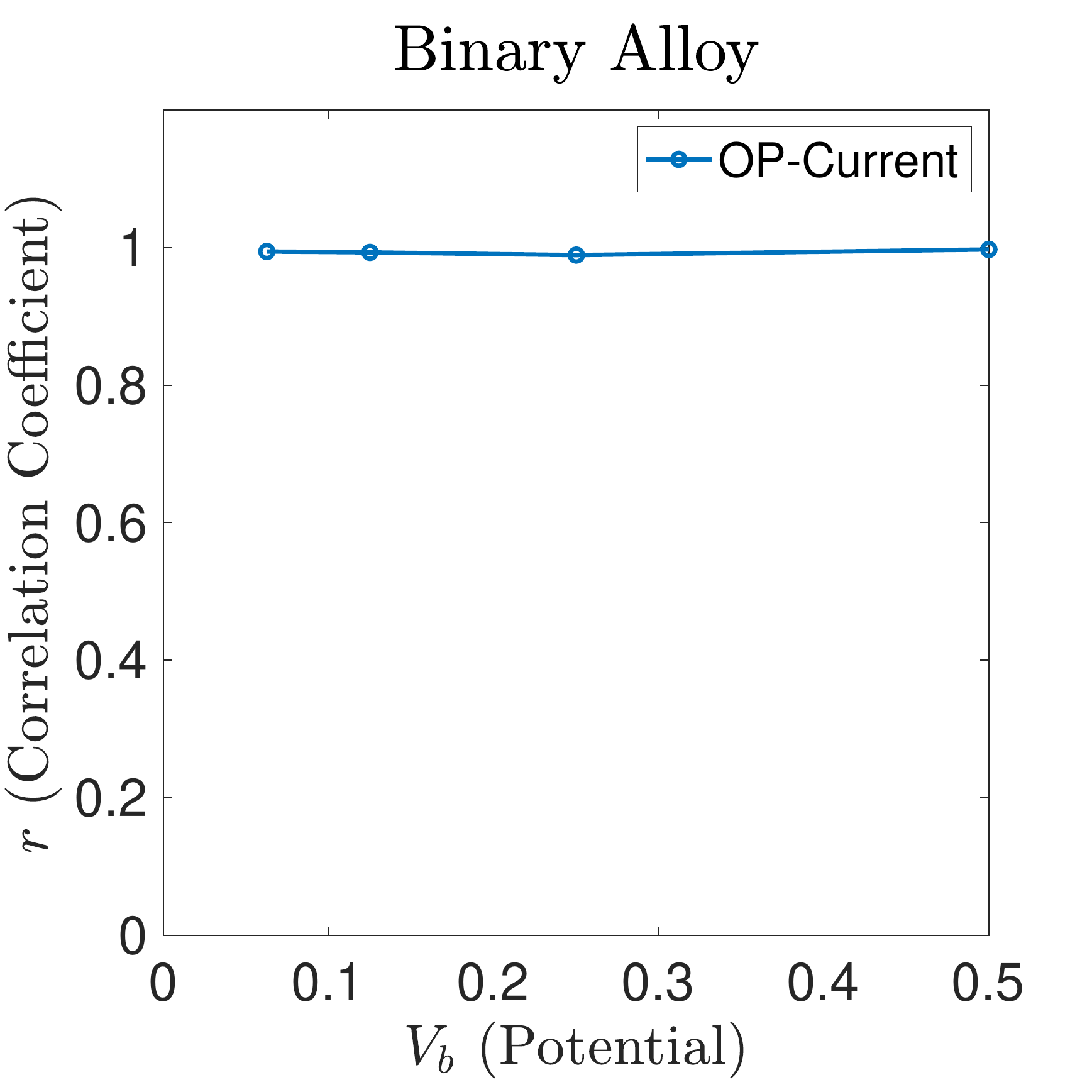}\quad
	\includegraphics[width=.3\textwidth]{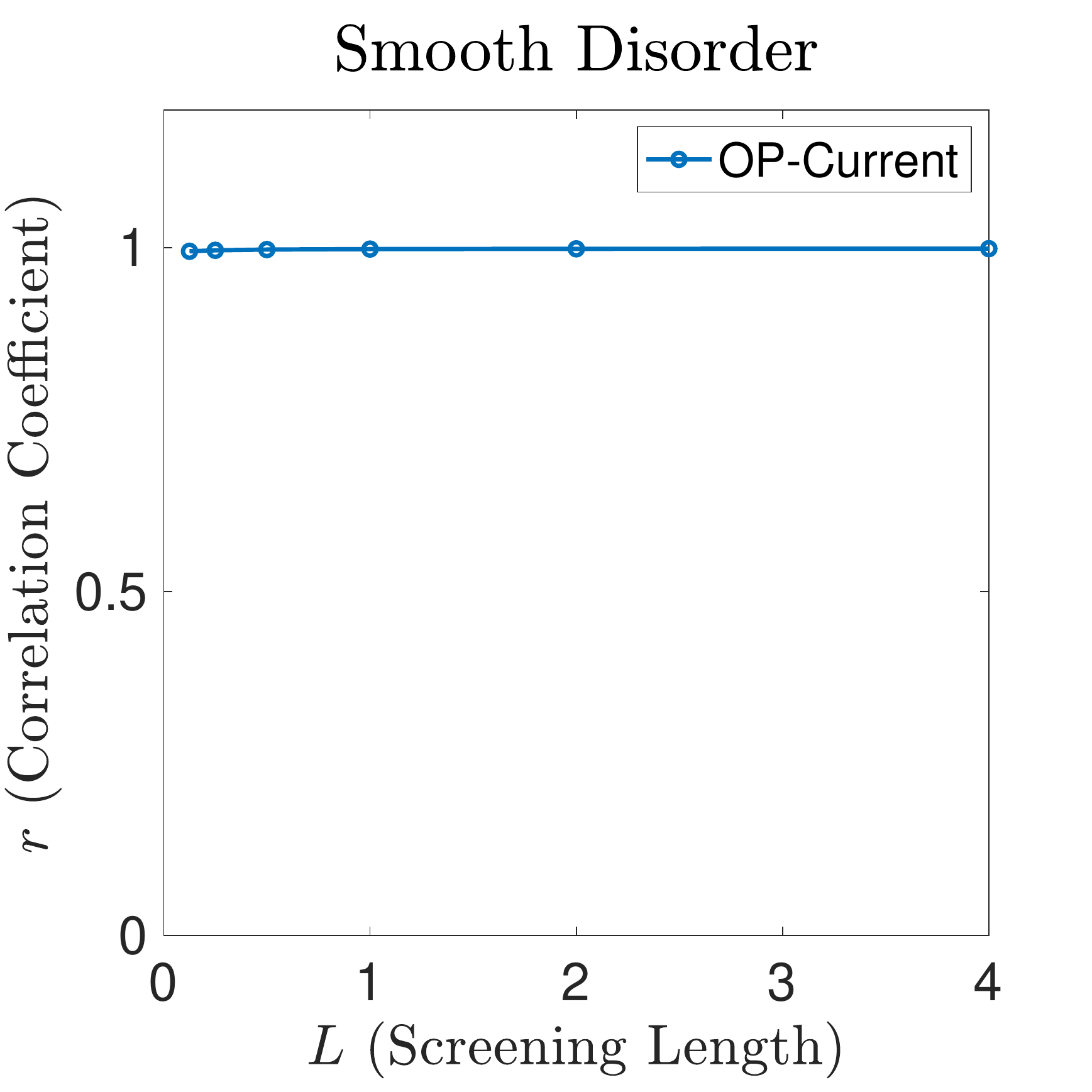}
	\caption{Correlation coefficient $r$ between the order parameter and $I_c$ for three different disorder types (weak pointlike disorder, binary-alloy disorder, and smooth disorder), shown as a function of  disorder parameters (impurity concentration $p$ for weak pointlike scatterers, impurity strength $V_b$ for binary-alloy disorder, and screening length $L$ for smooth disorder.) }
	\label{fig:gapcurrentcoefficients}
\end{figure*}

In this appendix, we discuss the correlation coefficient $r$ between the $d$-wave order parameter and $I_c$. We had earlier alluded to the fact that $ r \approx 1$ for all disorder types we have considered. The very close similarity between the aforementioned quantities had already been seen by Graham and Morr in a single-impurity context.\cite{graham2019josephson} For $s$-wave superconductors, Graham and Morr had also noted the nearly identical spatial dependence of these two quantities in the presence of various types of impurities.\cite{graham2017imaging} Here we show a number of explicit examples demonstrating the robustness of the correlation across different disorder types
	
In Figs.~\ref{fig:pointlikeopvscurrent},~\ref{fig:binaryalloyopvscurrent}, and~\ref{fig:smoothopvscurrent}, we show three different quantities for three different types of disorder. The first two plots are of the $d$-wave order parameter (Figs.~\ref{fig:pointlikeopvscurrent}a,~\ref{fig:binaryalloyopvscurrent}a, and~\ref{fig:smoothopvscurrent}a) and $I_c$ (Figs.~\ref{fig:pointlikeopvscurrent}b,~\ref{fig:binaryalloyopvscurrent}b, and~\ref{fig:smoothopvscurrent}b), while the last set of plots is for the normalized difference between the order parameter and the critical current---\emph{i.e.}, $\Delta(\mathbf{r}))/\overline{\Delta} - I_c(\mathbf{r})/\overline{I_c})$---which we use to highlight differences between the two quantities. The disorder types used are the same ones we had already shown in Figs.~\ref{fig:pointlikecombinedplot},~\ref{fig:binaryalloycombinedplot}, and~\ref{fig:smoothcombinedplot} (weak pointlike disorder with $p = 20\%$, binary-alloy disorder with $V_b = 0.250$, and smooth disorder with $L = 2$, respectively). It can be seen that the $d$-wave order parameter and $I_c$ look almost identical to each other, regardless of the disorder type used. There are differences between these two quantities, as can be seen in Figs.~\ref{fig:pointlikeopvscurrent}c,~\ref{fig:binaryalloyopvscurrent}c, and~\ref{fig:smoothopvscurrent}c, but the normalized difference is generally very small and is at most of the order of a few percent. In Fig.~\ref{fig:gapcurrenthistograms} we show two-dimensional histograms of the $d$-wave order parameter and $I_c$ for the aforementioned three types of disoder. It can be seen that the two quantities track each other very closely, with almost no deviation from the linear trend, regardless of the disorder type present. The correlation coefficients are all extremely close to 1.

We show the correlation coefficient $r$ between the order parameter and $I_c$  as a function of disorder parameters discussed in detail in the main text (\emph{i.e.}, impurity concentration for weak pointlike scatterers, impurity strength for binary-alloy disorder, and screening length for smooth disorder) in Fig.~\ref{fig:gapcurrentcoefficients}. Here we repeat the presentation of the correlation coefficients previously shown in Figs.~\ref{fig:pointcorr},~\ref{fig:binaryalloycorr}, and~\ref{fig:smoothcorr}. It can be seen across the three plots that $r \approx 1$, regardless of the disorder parameter---a much stronger correlation than any that between any other pairs of quantities. These results make clear that $I_c$ is an extremely good measure of the $d$-wave order parameter, regardless of the type or amount of disorder present in the superconductor. 


\bibliography{paper_jsts_disorder}

\end{document}